\documentclass[prl,letterpaper,twocolumn,showpacs,floatfix,superscriptaddress,amsmath,amsfonts,amssymb,preprintnumbers,citeautoscript]{revtex4-2}

\usepackage{CJK}
\usepackage{times}
\usepackage{graphicx}
\usepackage{amsmath}
\usepackage{amssymb}
\usepackage[colorlinks,linkcolor=blue,citecolor=blue,urlcolor=blue,hyperindex,pdfstartview=FitH,plainpages=false]{hyperref}
\usepackage{float}
\usepackage{lineno}

\newcommand{\Ts}{1$T$-TiSe$_2$}
\newcommand{\mJ}{mJ/cm$^2$}

\begin{document}

\begin{CJK*}{GBK}{}

\title{Ultrafast Switching from the Charge Density Wave Phase to a Metastable Metallic State in \Ts}
\author{Shaofeng Duan}
\affiliation{Key Laboratory of Artificial Structures and Quantum Control (Ministry of Education), School of Physics and Astronomy, Shanghai Jiao Tong University, Shanghai 200240, China}
\author{Wei Xia}
\affiliation{School of Physical Science and Technology, ShanghaiTech University, Shanghai 201210, China}
\author{Chaozhi Huang}
\author{Shichong Wang}
\author{Lingxiao Gu}
\author{Haoran Liu}
\affiliation{Key Laboratory of Artificial Structures and Quantum Control (Ministry of Education),  School of Physics and Astronomy, Shanghai Jiao Tong University, Shanghai 200240, China}

\author{Dao Xiang}
\affiliation{Key Laboratory for Laser Plasmas (Ministry of Education), School of Physics and Astronomy, Shanghai Jiao Tong University, Shanghai 200240, China}
\affiliation{Tsung-Dao Lee Institute, Shanghai Jiao Tong University, Shanghai 200240, China}
\author{Dong Qian}
\affiliation{Key Laboratory of Artificial Structures and Quantum Control (Ministry of Education), School of Physics and Astronomy, Shanghai Jiao Tong University, Shanghai 200240, China}
\affiliation{Tsung-Dao Lee Institute, Shanghai Jiao Tong University, Shanghai 200240, China}
\affiliation{Collaborative Innovation Center of Advanced Microstructures, Nanjing University, Nanjing 210093, China}
\author{Yanfeng Guo}
\affiliation{School of Physical Science and Technology, ShanghaiTech University, Shanghai, China}
\author{Wentao Zhang}
\email{wentaozhang@sjtu.edu.cn}
\affiliation{Key Laboratory of Artificial Structures and Quantum Control (Ministry of Education), School of Physics and Astronomy, Shanghai Jiao Tong University, Shanghai 200240, China}
\affiliation{Collaborative Innovation Center of Advanced Microstructures, Nanjing University, Nanjing 210093, China}
\date {\today}
\begin{abstract}

The ultrafast electronic structures of the charge density wave material \Ts~were investigated by high-resolution time- and angle-resolved photoemission spectroscopy. We found that the quasiparticle populations drove ultrafast electronic phase transitions in \Ts~within 100 fs after photoexcitation, and a metastable metallic state, which was significantly different from the equilibrium normal phase, was evidenced far below the charge density wave transition temperature. Detailed time- and pump-fluence-dependent experiments revealed that the photoinduced metastable metallic state was a result of the halted motion of the atoms through the coherent electron-phonon coupling process, and the lifetime of this state was prolonged to picoseconds with the highest pump fluence used in this study. Ultrafast electronic dynamics  were well captured by the time-dependent Ginzburg-Landau model. Our work demonstrates a mechanism for realizing novel electronic states by photoinducing coherent motion of atoms in the lattice.
 
\end{abstract}
\maketitle
\end{CJK*}

Manipulating the macroscopic properties of quantum materials via ultrafast photoexcitation can realize novel quantum phenomena that are not accessible in the thermal equilibrium condition and is also a fascinating field in condensed matter physics, such as the light-driven Floquet electronic states \cite{Wang2013}, optical amplification of the superconductivity or charge density wave (CDW) order \cite{Fausti2011, Mitrano2016, Wandel2022}, light-induced ferroelectricity \cite{Li2019, Nova2019}, photoinduced emergence of exotic metastable states \cite{Stojchevska2014, Morrison2014, Kogar2020}, and so on. 
Infrared optical irradiation with photon energy on the scale of 1 eV is mostly capable of interband electronic excitation, resulting in nonequilibrium quasiparticles near the Fermi energy through electron-electron and electron-phonon scattering processes in tens of femtoseconds \cite{Basov2011}.
A sudden change of carrier density would transiently enhance the screening of electronic correlations and induce displacive forces to modulate the equilibrium position of specific atoms in the lattice and generate the coherent phonon modes \cite{Giulietti2012}. 
The intertwined couplings among the orbital, spin, lattice, and charge degrees of freedom give rise to complex phases in quantum materials. Therefore, if such a photoinduced coherent motion of atoms could stop at metastable positions away from their equilibrium counterparts, the material would possibly enter into a novel metastable state different from its original physics. Such optical manipulation of quantum materials via halting the photoinduced coherent phonon mode has rarely been studied in real materials.

Optical-induced coherent phonon modes have been observed in multiple systems, such as the high-temperature superconductors \cite{Gerber2017, Yang2019a, Yang2022}, the CDW materials \cite{Maklar2021, Shi2019, Duan2021}, the topological insulators \cite{Sobota2014}, and other correlated systems \cite{Bretscher2021, Tang2020}. Among these states, CDW is a unique symmetry-broken state with periodic modulation of the lattice structure and charge density, mediated by electron-phonon or electron-electron interactions. The transition-metal dichalcogenide \Ts~is a typical CDW material and undergoes a commensurate CDW phase transition at 202 K ($T_c$),  accompanied by a 2$\times$2$\times$2 structural transition \cite{DiSalvo1976}, and the CDW is possibly a result of the hybrid Jahn-Teller effect and excitonic condensation \cite{Kogar2017, Porer2014, Cheng2022, Cercellier2007, Hedayat2019}. In addition, there are many novel quantum phases in \Ts, such as superconductivity by Cu doping \cite{Morosan2006} or applying pressure \cite{Kusmartseva2009, Joe2014, Li2016}, photoinduced gyrotropic electronic order \cite{Xu2020} and chiral CDW \cite {Wickramaratne2022}, and light-induced two-dimensional electronic states with potential energy gap opening \cite{Duan2021}. The intertwined interactions and rich phase diagrams make \Ts~a good platform for exploring photoinduced novel phases in a nonthermal way. Moreover, photoinduced coherent phonon modes have been identified in \Ts~by multiple time-resolved techniques \cite{Porer2014,Hedayat2019,Holy1977,Snow2003}. Thus, taking advantage of improved time and energy resolutions in time-resolved experiments, it is possible to realize novel phases in \Ts~by coherently tuning the motion of atoms in the lattice. 
 
In this Letter, we report the observation of a fully screened metastable metallic state right after the CDW order melting by time- and angle-resolved photoemission spectroscopy (TRARPES). Specially, we found that the CDW order was melted above a critical pump fluence of about 0.12 \mJ, and the two Se $4p_{x,y}$ valence bands shifted up toward the Fermi energy by about 120 meV, of which the value was almost fluence independent, exhibiting a plateau in the time-dependent energy shifts before the recovery of electronic correlations.  Moreover, the photoinduced metastable states lasted longer with higher pump fluence and persisted longer than 1 ps for the highest excitation fluence we studied. In addition, we found that the energy bands oscillated around the metastable position with an anharmonic frequency much higher than  the $A_\text{1g}$-CDW coherent phonon mode at low excitation fluence. Such phenomena were  absent above the CDW transition temperature. Combing with the numerical solution of a temporally dependent Ginzburg-Landau potential, we attributed the emergence of  the observed metastable state to the optical-induced coherent phonon mode and halted coherent motion of the related atoms in the lattice.

Ultrafast electronic structure measurements were performed with a home-built TRARPES system at a repetition rate of 500 kHz \cite{Yang2019}. In the measurements, infrared laser pulses with a wavelength centered at 700 nm ($h\nu$ = 1.77 eV) drove the system into nonequilibrium states, and then ultraviolet laser pulses with a wavelength centered at 205 nm ($h\nu$ = 6.05 eV) photoemit electrons. Time resolution was achieved by varying the delay between the pump and probe pulses. The overall energy and time resolutions were optimized to 16 meV and 113 fs, respectively \cite{Huang2022}. High-quality single crystals of 1T-TiSe$_2$ were grown by chemical vapor transport with an iodine transport agent at grown temperature 650 $\rm{}^{\circ }C$, and the sample was cleaved in an ultrahigh vacuum condition with a base pressure better than 3$\times$ 10$^{-11}$ Torr. 

\begin{figure}
\centering\includegraphics[width = 1\columnwidth] {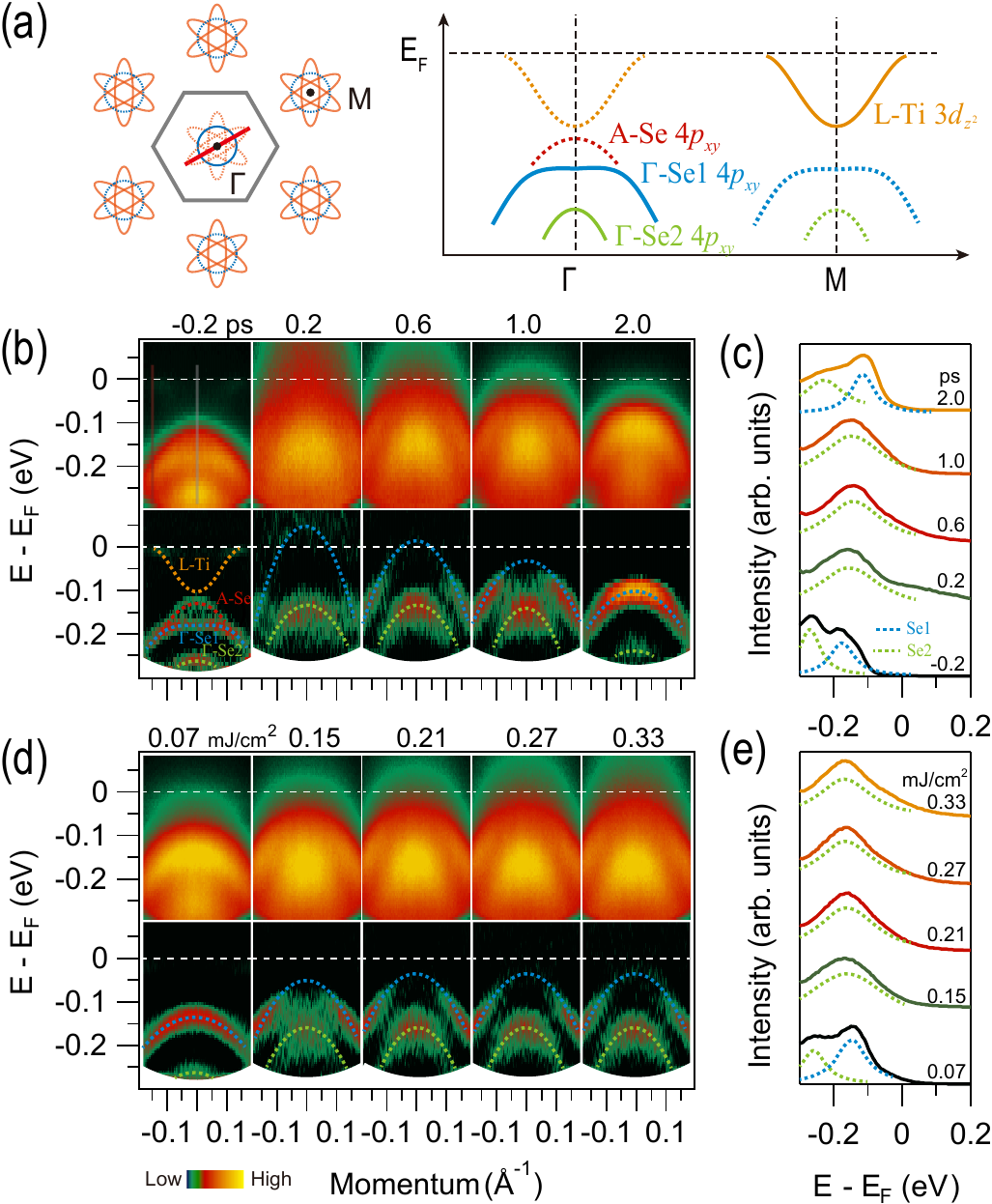}
\caption{
TRARPES spectra measured on \Ts~ near the Brillouin zone center along the $\Gamma $-$M$ at an equilibrium temperature of 4 K. (a) Schematics of the electronic structure in the charge density wave state. (b) TRARPES spectra measured at the delay times of $-$0.2, 0.2 0.6, 1.0, and 2.0 ps (top) with the pump fluence of 0.356 \mJ~ and the corresponding second-derivative images (bottom). The gray and dark red lines denote the cuts at $\Gamma$ and -0.16 $\mathring{A}^{-1}$, respectively. (c) Energy distribution curves (EDCs) at $\Gamma$ from (b). (d) TRARPES spectra at the delay time of 0.5 ps with pump fluences of 0.07, 0.15, 0.21, 0.27, and 0.33 \mJ~ (top) and the corresponding second-derivative images (bottom). (e) EDCs at $\Gamma$ point from (d). The dotted curves in (c) and (e) are the Lorentzian fitting curves to capture the peak positions of the Se1 and Se2 bands. }
\label{Fig1}
\end{figure}

At an equilibrium temperature of 4 K, far below the CDW transition temperature, the equilibrium photoemission spectra at a delay time of $-$0.2 ps showed four bands near the Brillouin-zone center, including the Ti 3$d_{z^2}$ conduction band folded from the $L$ point, the Se 4$p_{x,y}$ band folded from the $A$ point, and a pair of Se 4$p_{x,y}$ valence bands derived from the $\Gamma$ point (Figs. \ref{Fig1}(a) and \ref{Fig1}(b) and Supplemental Material, Discussion No. 1 \cite{SM}).
The apparent flat dispersion of the Se1 band was a signature of the mixing of the $L$-derived Ti 3$d_{z^2}$ conduction band and the Se1 valence band due to the strong electron-hole interaction \cite{Pillo2000, Cercellier2007, Watson2019}. The Se2 band was also strongly coupled to the CDW order and exhibited significant downshifts through the CDW phase transition \cite{Watson2019}.

After photoexciting the sample with a fluence of 0.356 \mJ, higher than the melting threshold of excitonic correlation and CDW order \cite {Hedayat2019, Duan2021}, the band top of Se1 could not be well defined within about 1 ps, which is possibly a result of the strong excitonic fluctuation and photoemission matrix element effect \cite{Cercellier2007}.  However, the Se1 band crossed the Fermi energy before about 1 ps after melting the CDW order, suggesting a photoinduced transient metallic state within the time window.
Nevertheless, the energy shift of the Se1 band was resolved at the momentum away from the $\Gamma$ point, since the CDW order fluctuation contributed less to the electronic structure at higher binding energies.
The energy band at the binding energy of about -0.16 eV and at the zero momentum was attributed to the Se2 valence band within 1 ps due to the vanished intensity of the Se1 band top (Supplemental Material, Discussion No. 2 \cite{SM}).
In addition, the Se2 band shifted up by about 120 meV after photoexcitation, and interestingly, such band shift was nearly time independent within 1.0 ps, which was further evidenced by the time-dependent energy distribution curves (EDCs) at $\Gamma$ in Fig. \ref{Fig1}(c) (Supplemental Material, Discussion No. 3 \cite{SM}). Both the Se1 and Se2 bands were partially restored after 2.0 ps. The pump-induced energy shift of the Se2 band was not dependent on the excitation density  after CDW order melting, as shown in Figs. \ref{Fig1}(d) and \ref{Fig1}(e). The time- and pump-fluence-independent band shifts observed before the build-in of the electronic correlations suggest the system entered into a metastable metallic state after CDW order melting.

\begin{figure*}
\centering\includegraphics[width = 2\columnwidth] {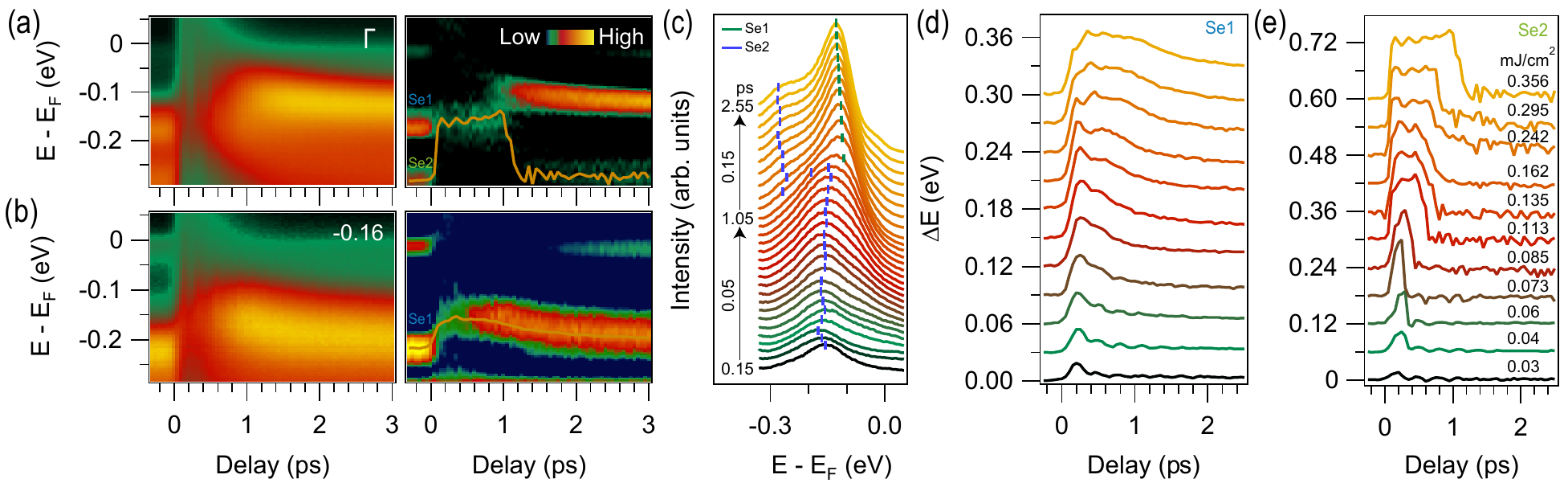}
\caption{
Ultrafast electronic dynamics in the CDW state. (a) Photoemission intensity as a function of binding energy and delay time at $\Gamma$ with the pump fluence of 0.356 \mJ~and the corresponding second-derivative image (right).  (b) Same as (a) but measured at the momentum of -0.16 $\mathring{A}^{-1}$. The orange solid lines in (a) and (b) are the peak positions of the Se2 and Se1 bands determined from Lorentzian fitting. (c) EDCs from (a) at selected delay times.
The green and blue bars indicate the peak position of the Se1 and Se2 $4p_{x,y}$ bands, respectively. (d) Time-dependent energy shifts of the  Se1 $4p_{x,y}$ band at the momentum of -0.16 $\mathring{A}^{-1}$ at selected pump fluences. (e) Same as (d) but of the  Se2 $4p_{x,y}$ band at $\Gamma$ point. We note that the positions of the Se2 band cannot be well defined right after the time delay at the end of the plateau before the recovery of  well-defined EDC peaks of the Se2 band.}
\label{Fig2}
\end{figure*}

To investigate the physics of this metastable state, we focused on the ultrafast dynamics of the Se2 band at $\Gamma$ and the Se1 band at the momentum of -0.16 $\mathring{A}^{-1}$, at which of the momentum the Se1 band was clearly resolved in the entire delay time range. Right after strong optical excitation with a density of 0.356 \mJ, the Se2 band shifted upward by about 120 meV toward the Fermi energy, exhibiting a plateau feature within 1 ps (Fig. \ref{Fig2}(a)). The recovery of the coherent quasiparticle peaks of the Se1 band after about 1 ps suggests that the long-range CDW was gradually restored, and at the same time, the Se2 band almost jumped back to its original energy within 200 fs. In the time-dependent spectra at the momentum of -0.16 $\mathring{A}^{-1}$, the Se1 band also showed the signature of a plateau before about 1 ps after photoexcitation and gradually moved back to its equilibrium energy as the CDW order recovered (Fig. \ref{Fig2}(b)). Such plateau features in the Se1 and Se2 bands suggest the system entered into a metastable state after CDW order melting. The lifetime of this metastable state is enhanced as the pump fluence increased (Figs. \ref{Fig2}(d) and \ref{Fig2}(e) and Supplemental Material, FIGs. 3(c)-3(e) and  Discussion No. 4 \cite{SM}). This metastable state  is far different from metastable states in other CDW materials such as 1$T$-TaSe$_2$, in which the photoinduced band shifts recover quickly after photoexcitations and the claimed metastable state occurs at a much longer delay time \cite{Shi2019}. Previously, a transient plateau feature of the band energy was also identified in K$_{0.3}$MoO$_{0.3}$ with the lifetime limited by the period of the CDW amplitude mode \cite{Yang2020}, which is different from the  plateau with tunable lifetime observed herein.

In addition, the time-dependent photoemission spectra showed obvious oscillating features in both the Se1 and Se2 bands within 1 ps (Figs. \ref{Fig2}(a) and \ref{Fig2}(b)), as shown in the EDCs at $\Gamma$ as a function of delay time (Fig. \ref{Fig2}(c)). Previous study showed that such oscillation in the plateau is at a higher frequency than that of the photoinduced coherent $A_{1g}$-CDW phonons at low pump fluence and even in an anharmonic behavior (Figs. \ref{Fig2}(d)-\ref{Fig2}(e) and Supplemental Material, Discussions No.  4 and No. 5 \cite{SM}) \cite{Duan2021}, suggesting a photoinduced ultrafast structural transition in the time window of the plateau. Ultrafast electron diffraction measurement has shown that the CDW structure distortion is recovered to its normal phase right after strong photoexcitations, but the temperature of the lattice is still far below the transition temperature \cite{Cheng2022}.

Furthermore, the photoinduced metastable state observed in this study could be only realized in the CDW state. In the normal state, photoexcitation only served to reach a transiently high electronic temperature and excited electrons to the unoccupied states, which did not launch the coherent excitation of the order parameter. After photoexcitation, the valence bands shifted up and recovered quickly to their equilibrium states with no coherent oscillation and plateau feature as identified in the CDW state (Supplemental Material, FIGs. 5(c)-5(e) and Discussion No. 6 \cite{SM}). The absence of such photoinduced metastable states in the equilibrium state suggests that collective excitation of coherent phonon is essential for driving the metastable state far from equilibrium.

\begin{figure}
\centering\includegraphics[width = 1\columnwidth] {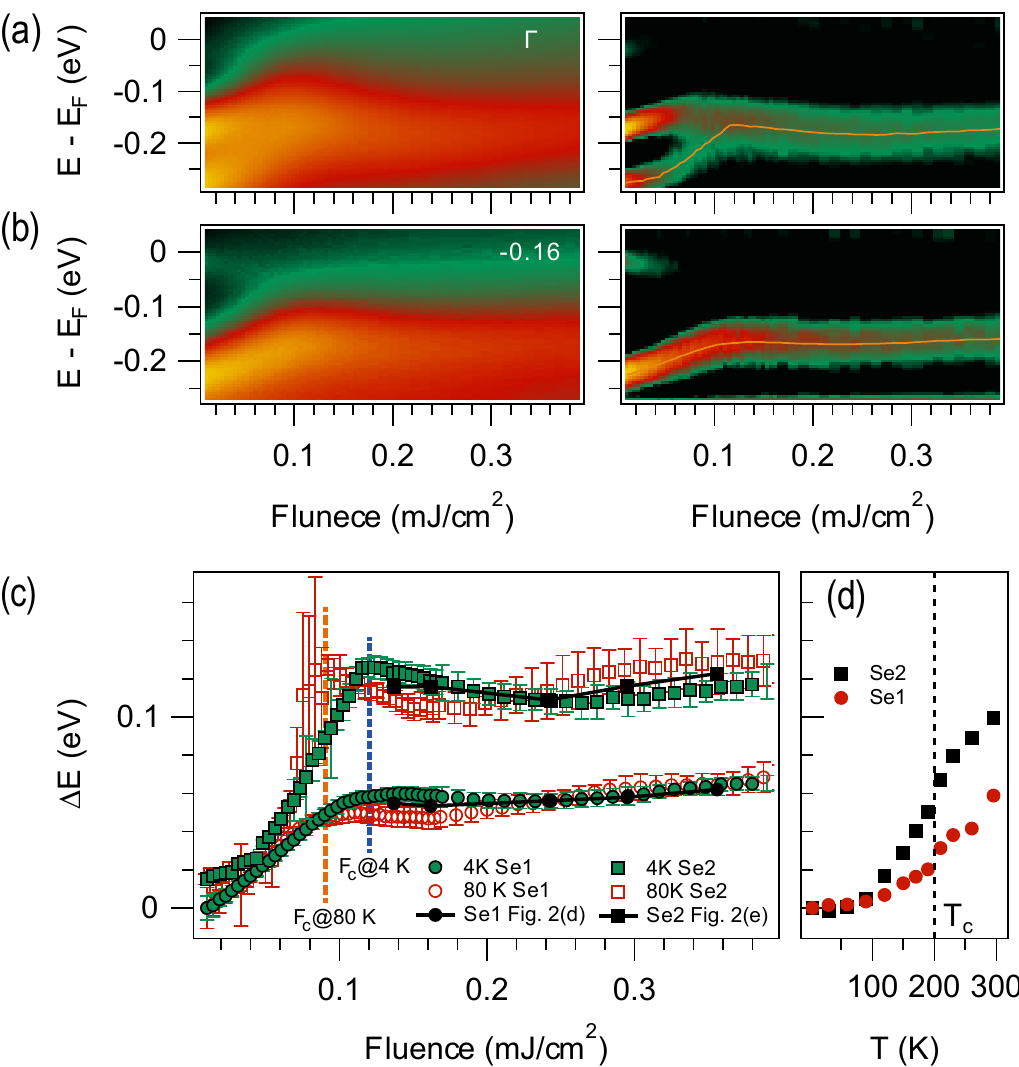}
\caption{
Fluence- and temperature-dependent TRARPES spectra. (a) Fluence-dependent photoemission spectra at $\Gamma$ point and delay time 0.3 ps (left) and the corresponding second-derivative image (right). (b) Same as (a) but measured at the momentum -0.16 $\mathring{A}^{-1}$. (c) Fluence-dependent energy shifts of the Se1 and Se2 bands at equilibrium temperatures 4.5 and 80 K. (d)  Temperature-dependent energy shifts of the Se1 and Se2 bands.}
\label{Fig3}
\end{figure}

Such a photoinduced metastable state right after CDW order melting is quite different from the normal equilibrium state above the transition temperature. By finely tuning the pump fluence at the delay time of 0.3 ps (in the plateau), both the Se1 and Se2 bands shifted up monotonously until reaching a critical pump fluence ($F_c$) of about 0.12 \mJ, which is the threshold of quenching the CDW order. Above the critical fluence, the shifts of the Se1 and Se2 bands saturated at about 60 and 120 meV, respectively (Figs. \ref{Fig3}(a)-\ref{Fig3}(c)). Specially, the estimated energy shifts of the Se1 band at the $\Gamma$ point was about 120 meV. The slight downshift above $F_c$ was a result of the varying anharmonic oscillation frequency at high pump fluence.
These observations were also repeatable at 80 K but with a lower critical pump fluence of about 0.09 \mJ. Above the CDW transition temperature the equilibrium energies of the Se1 and Se2 bands did not saturate and still generally shifted upward largely with increasing temperature. Such shifts of the energy band above the transition temperature are possibly a result of the continuing melting of the in-plane two-dimensional CDW and the possible temperature-induced chemical potential shift \cite{Cheng2022, Chen2016, Monney2010}. The above observations suggest that strong ultrafast photoexcitations have fully quenched the CDW order in a short timescale with the lattice temperature still far below the $T_c$.

After pumping the sample at 300 K, there was no plateau feature and no similar anharmonic oscillation in the time-dependent energy shifts (Supplemental Material, FIGs. 5(c)-5(e) and Discussion No. 6 \cite{SM}), and the total energy shifts (the sum of maximum photoinduced energy shifts at 300 K and the equilibrium energy differences between 4 and 300 K) were about 35\% larger than the energy shifts of the photoinduced metastable states at low temperature. 
These observations suggest that the photoinduced metastable states do not simply hold the similarity as the normal electronic structure, and may result from novel pathways, such as the ultrafast modification of the free energy potential, the ultrafast structural transition, and the strong modification of the charge screening effect due to photoinduced nonequilibrium charge carriers. 

\begin{figure*}
\centering\includegraphics[width = 2\columnwidth] {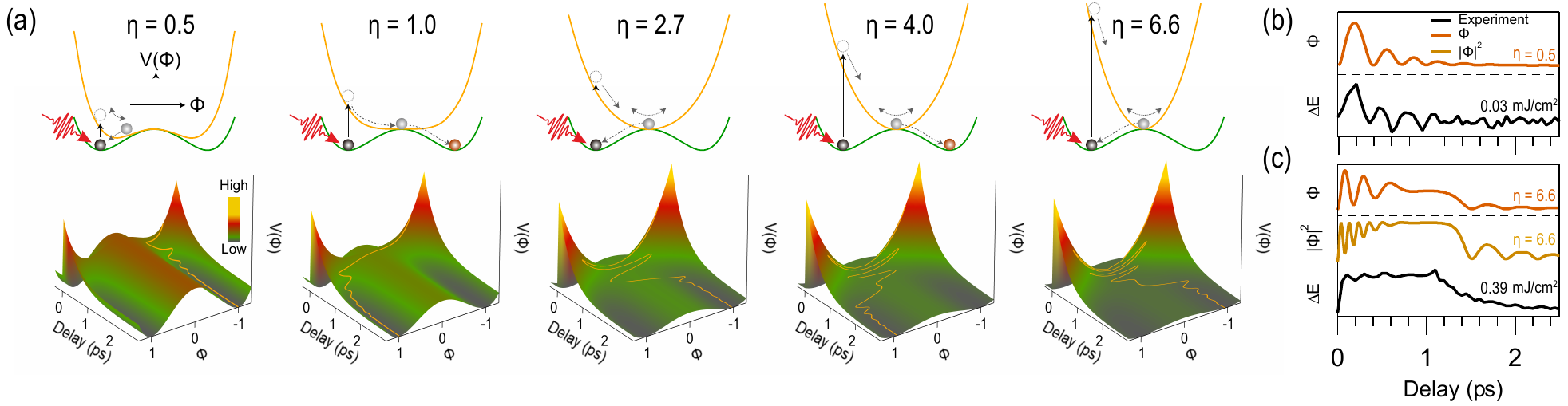}
\caption{
Numerical solution of the order parameter dynamics from the time-dependent Ginzburg-Landau model. (a) Energy potentials before and right after photoexcitation (top) and transient energy potential as a function of order parameter and delay time and the order parameter dynamics (bottom) for selected pump fluences $\eta$ = 0.5, 1.0, 2.7, 4.0, and 6.6. (b) and (c) Experimental and simulated order parameter dynamics at low and high pump fluence respectively. }
\label{Fig4}
\end{figure*}

To understand the underlying physics of this photoinduced metastable state, we adopted the temporally dependent double-well Ginzburg-Landau model to simulate the electronic states right after photoexcitation \cite{Trigo2021, Yusupov2010}. 
The time-dependent potential on the sample surface with the order parameter $\Phi$ is
\begin{equation}
	V\left( \Phi\right)  =\frac{1}{4} \left( \eta(t)  -1\right)  \Phi^{2} +\frac{1}{8} \Phi^{4} .
\label{eq1}
\end{equation}
Here, $\eta(t)=\eta e^{ -{t}/{\tau_{e} } }$ is the transient modification of the potential with the pump fluence $\eta$ after photoexcitation, and the nonequilibrium quasiparticle lifetime $\tau_{e}$ = 0.5 ps was estimated from experiments. A full time-dependent potential away from the sample surface is given in Ref. \cite{Duan2021}.
The system in the CDW ground state corresponds to the double-well potential with the order parameter $\Phi$ = $-$1. Numerical solutions of the order parameter as a function of delay time were obtained from the motion equation in Supplemental Material (Discussion No. 7 \cite{SM}).

At a very low pump fluence with $\eta$ = 0.5 in the simulation, the potential was weakly modified after photoexcitation and still maintained its double-well shape, which launched a damped oscillation with the CDW amplitude mode and then recovered to its equilibrium ground state ($\Phi$ = $-$1) within 2.5 ps (Fig. \ref{Fig4}(a)). As the pump fluence is increased with $\eta$ = 1.0, the $V(\Phi)$ was transiently modified after photoexcitation and showed only one minimum (single-well shape) with $\Phi=0$, suggesting the full suppression of the CDW order.  Interestingly, the order parameter stepped over the double-well barrier near $\Phi=0$ one time and eventually fell at $\Phi$ = 1, suggesting the photoinduced phase inversion at this pump fluence.  The order parameter of $\Phi$ = 1 is a fully inverted state, which shares the same symmetry as the original state ($\Phi$ = $-$1), characterized by order parameters with equal magnitudes but opposite phases. As the pump fluence  increased to $\eta$ = 2.7, the photoinduced single-well potential became steeper and the order parameter stepped over the barrier four times,  fell at $\Phi$ = $-$1, and then recovered to its original phase. At the pump fluences of $\eta$ = 4 and 6.6, the order parameter stepped over the barrier five times {(recovering to the inverted phase $\Phi$ = 1) and six times (recovering to the initial phase $\Phi$ = $-$1), respectively.
At longer delay times, the sectional inversion of order parameters within the sample will induce interfaces existing parallel to the sample surface at the interval between the inverted phase ($\Phi$ = $+$1) and the original phase ($\Phi$ = $-$1) \cite{Duan2021}, and the final states at specific pump fluence can be roughly estimated from the detailed fluence-dependent electronic structure measurements (Supplemental Material, Discussion No. 8 \cite{SM}).

Next, we investigated the physics of the observed metastable states at the delay time right after photoexcitation.
At low pump fluence, we found that the simulated time-dependent order parameter repeated the experimental energy band shift as a function of delay time precisely (Fig. \ref{Fig4}(b)). At the high pump fluence of $\eta$ = 6.6, the order parameter oscillated anharmonically around a new position at which  $\Phi$ was equal to 0, exhibiting a plateau similar to the experimental data at high pump fluence (Fig. \ref{Fig4}(c)), and such behaviors were also evidenced in the square of the time-dependent order parameter.
 The accurate simulation clearly suggests that the photoinduced metastable state tends to be a novel phase with the dynamical CDW lattice distortion stabilized around $\Phi$ = 0 due to the coherent motion of the atoms associated with the $A_{1g}$-CDW phonon mode. Recent theoretical and experimental studies on nonthermal disturbed CDW order in \Ts~ have also confirmed that optical excitation of the electronic state could drive atoms to move toward the undistorted ground state \cite{heinrich, Lian2020}.

We finally addressed the question of to what extent the screening by the photoinduced charge carriers affects the properties of \Ts~on the picosecond timescale. With the pump fluence above $F_c$ the shifts of both Se1 and Se2 bands underwent sudden transitions within 100 fs (Figs.~\ref{Fig2}(d) and \ref{Fig2}(e)), of which the value was nearly fluence independent. The sudden energy shifts after photoexcitation is indicative of transiently melting the exciton and CDW order in \Ts. Multiple experiments have demonstrated that photoexcited charge carriers could enhance the Coulomb screening and quench the excitonic order on ultrafast timescale \cite{Mor2017, Okazaki2018, Porer2014}.
Such photoinduced charge carrier screening effects can be as fast as 20 fs in an experiment with better time resolution \cite{Rohwer2011}. In addition, the plateau behavior in the band shifts as a function of delay time, and the saturated band shifts above $F_c$ indicate that the photoinduced charge carriers screened the electron-electron interaction completely, and the observed nonequilibrium electronic states were the bare dispersions for the undistorted lattice at low temperature. Considering the charge carrier screen effect, the longer lifetime of the photoinduced metastable metallic state is consistent with the observation in our previous study that the nonequilibrium quasiparticle recovery time is longer at higher pump fluence \cite{Duan2021}.

In conclusion, we used TRARPES to investigate the time- and fluence-dependent electronic structure dynamics in \Ts, revealing a fully screened metallic electronic state after  CDW order melting. The lifetime of this photoinduced metastable state was tunable by varying the pump fluence and could be longer than 1 ps with strong excitation, and it can be potentially extended to much longer by enhancing the photoexcitation fluence. Through time-dependent Ginzburg-Landau simulation, we demonstrated that the emergence of this metastable  state is due to the coherent excitation of the amplitude mode and the photoinduced transient modification of the double-well potential in the CDW phase. This new state enriches the phase diagram of \Ts~ and paves the way for developing similar methods to manipulate and control other ordered phases with ultrafast light.

\begin{acknowledgements}
W. T. Z. acknowledges support from the National Key R\&D Program of China (Grants No. 2021YFA1400202 and No. 2021YFA1401800) and National Natural Science Foundation of China (Grants No. 12141404 and No. 11974243) and Natural Science Foundation of Shanghai (Grants No. 22ZR1479700 and No. 23XD1422200). S. F. D. acknowledges support from the China Postdoctoral Science Foundation (Grant No. 2022M722108). Y. F. G. acknowledges the support from the National Natural Science Foundation of China (No. 92065201) and the Double First-Class Initiative Fund of ShanghaiTech University. D.X. acknowledges support from the National Natural Science Foundation of China (Grant No. 11925505). D.Q. acknowledges support from the National Key R\&D Program of China (Grants No. 2022YFA1402400 and No. 2021YFA1400100) and National Natural Science Foundation of China (Grant No. 12074248).
\end{acknowledgements}


\begin{thebibliography}{48}%
\makeatletter
\providecommand \@ifxundefined [1]{%
 \@ifx{#1\undefined}
}%
\providecommand \@ifnum [1]{%
 \ifnum #1\expandafter \@firstoftwo
 \else \expandafter \@secondoftwo
 \fi
}%
\providecommand \@ifx [1]{%
 \ifx #1\expandafter \@firstoftwo
 \else \expandafter \@secondoftwo
 \fi
}%
\providecommand \natexlab [1]{#1}%
\providecommand \enquote  [1]{``#1''}%
\providecommand \bibnamefont  [1]{#1}%
\providecommand \bibfnamefont [1]{#1}%
\providecommand \citenamefont [1]{#1}%
\providecommand \href@noop [0]{\@secondoftwo}%
\providecommand \href [0]{\begingroup \@sanitize@url \@href}%
\providecommand \@href[1]{\@@startlink{#1}\@@href}%
\providecommand \@@href[1]{\endgroup#1\@@endlink}%
\providecommand \@sanitize@url [0]{\catcode `\\12\catcode `\$12\catcode
  `\&12\catcode `\#12\catcode `\^12\catcode `\_12\catcode `\%12\relax}%
\providecommand \@@startlink[1]{}%
\providecommand \@@endlink[0]{}%
\providecommand \url  [0]{\begingroup\@sanitize@url \@url }%
\providecommand \@url [1]{\endgroup\@href {#1}{\urlprefix }}%
\providecommand \urlprefix  [0]{URL }%
\providecommand \Eprint [0]{\href }%
\providecommand \doibase [0]{https://doi.org/}%
\providecommand \selectlanguage [0]{\@gobble}%
\providecommand \bibinfo  [0]{\@secondoftwo}%
\providecommand \bibfield  [0]{\@secondoftwo}%
\providecommand \translation [1]{[#1]}%
\providecommand \BibitemOpen [0]{}%
\providecommand \bibitemStop [0]{}%
\providecommand \bibitemNoStop [0]{.\EOS\space}%
\providecommand \EOS [0]{\spacefactor3000\relax}%
\providecommand \BibitemShut  [1]{\csname bibitem#1\endcsname}%
\let\auto@bib@innerbib\@empty
\bibitem [{\citenamefont {Wang}\ \emph {et~al.}(2013)\citenamefont {Wang},
  \citenamefont {Steinberg}, \citenamefont {Jarillo-Herrero},\ and\
  \citenamefont {Gedik}}]{Wang2013}%
  \BibitemOpen
  \bibfield  {author} {\bibinfo {author} {\bibfnamefont {Y.~H.}\ \bibnamefont
  {Wang}}, \bibinfo {author} {\bibfnamefont {H.}~\bibnamefont {Steinberg}},
  \bibinfo {author} {\bibfnamefont {P.}~\bibnamefont {Jarillo-Herrero}},\ and\
  \bibinfo {author} {\bibfnamefont {N.}~\bibnamefont {Gedik}},\ }\href
  {https://doi.org/10.1126/science.1239834} {\bibfield  {journal} {\bibinfo
  {journal} {Science}\ }\textbf {\bibinfo {volume} {342}},\ \bibinfo {pages}
  {453} (\bibinfo {year} {2013})}\BibitemShut {NoStop}%
\bibitem [{\citenamefont {Fausti}\ \emph {et~al.}(2011)\citenamefont {Fausti},
  \citenamefont {Tobey}, \citenamefont {Dean}, \citenamefont {Kaiser},
  \citenamefont {Dienst}, \citenamefont {Hoffmann}, \citenamefont {Pyon},
  \citenamefont {Takayama}, \citenamefont {Takagi},\ and\ \citenamefont
  {Cavalleri}}]{Fausti2011}%
  \BibitemOpen
  \bibfield  {author} {\bibinfo {author} {\bibfnamefont {D.}~\bibnamefont
  {Fausti}}, \bibinfo {author} {\bibfnamefont {R.~I.}\ \bibnamefont {Tobey}},
  \bibinfo {author} {\bibfnamefont {N.}~\bibnamefont {Dean}}, \bibinfo {author}
  {\bibfnamefont {S.}~\bibnamefont {Kaiser}}, \bibinfo {author} {\bibfnamefont
  {A.}~\bibnamefont {Dienst}}, \bibinfo {author} {\bibfnamefont {M.~C.}\
  \bibnamefont {Hoffmann}}, \bibinfo {author} {\bibfnamefont {S.}~\bibnamefont
  {Pyon}}, \bibinfo {author} {\bibfnamefont {T.}~\bibnamefont {Takayama}},
  \bibinfo {author} {\bibfnamefont {H.}~\bibnamefont {Takagi}},\ and\ \bibinfo
  {author} {\bibfnamefont {A.}~\bibnamefont {Cavalleri}},\ }\href
  {https://doi.org/10.1126/science.1197294} {\bibfield  {journal} {\bibinfo
  {journal} {Science}\ }\textbf {\bibinfo {volume} {331}},\ \bibinfo {pages}
  {189} (\bibinfo {year} {2011})}\BibitemShut {NoStop}%
\bibitem [{\citenamefont {Mitrano}\ \emph {et~al.}(2016)\citenamefont
  {Mitrano}, \citenamefont {Cantaluppi}, \citenamefont {Nicoletti},
  \citenamefont {Kaiser}, \citenamefont {Perucchi}, \citenamefont {Lupi},
  \citenamefont {Di~Pietro}, \citenamefont {Pontiroli}, \citenamefont
  {Ricc$\rm\grave{o}~$}, \citenamefont {Clark}, \citenamefont {Jaksch},\ and\
  \citenamefont {Cavalleri}}]{Mitrano2016}%
  \BibitemOpen
  \bibfield  {author} {\bibinfo {author} {\bibfnamefont {M.}~\bibnamefont
  {Mitrano}}, \bibinfo {author} {\bibfnamefont {A.}~\bibnamefont {Cantaluppi}},
  \bibinfo {author} {\bibfnamefont {D.}~\bibnamefont {Nicoletti}}, \bibinfo
  {author} {\bibfnamefont {S.}~\bibnamefont {Kaiser}}, \bibinfo {author}
  {\bibfnamefont {A.}~\bibnamefont {Perucchi}}, \bibinfo {author}
  {\bibfnamefont {S.}~\bibnamefont {Lupi}}, \bibinfo {author} {\bibfnamefont
  {P.}~\bibnamefont {Di~Pietro}}, \bibinfo {author} {\bibfnamefont
  {D.}~\bibnamefont {Pontiroli}}, \bibinfo {author} {\bibfnamefont
  {M.}~\bibnamefont {Ricc$\rm\grave{o}~$}}, \bibinfo {author} {\bibfnamefont
  {S.~R.}\ \bibnamefont {Clark}}, \bibinfo {author} {\bibfnamefont
  {D.}~\bibnamefont {Jaksch}},\ and\ \bibinfo {author} {\bibfnamefont
  {A.}~\bibnamefont {Cavalleri}},\ }\href {https://doi.org/10.1038/nature16522}
  {\bibfield  {journal} {\bibinfo  {journal} {Nature}\ }\textbf {\bibinfo
  {volume} {530}},\ \bibinfo {pages} {461} (\bibinfo {year}
  {2016})}\BibitemShut {NoStop}%
\bibitem [{\citenamefont {Wandel}\ \emph {et~al.}(2022)\citenamefont {Wandel},
  \citenamefont {Boschini}, \citenamefont {da~Silva~Neto}, \citenamefont
  {Shen}, \citenamefont {Na}, \citenamefont {Zohar}, \citenamefont {Wang},
  \citenamefont {Welch}, \citenamefont {Seaberg}, \citenamefont {Koralek},
  \citenamefont {Dakovski}, \citenamefont {Hettel}, \citenamefont {Lin},
  \citenamefont {Moeller}, \citenamefont {Schlotter}, \citenamefont {Reid},
  \citenamefont {Minitti}, \citenamefont {Boyle}, \citenamefont {He},
  \citenamefont {Sutarto}, \citenamefont {Liang}, \citenamefont {Bonn},
  \citenamefont {Hardy}, \citenamefont {Kaindl}, \citenamefont {Hawthorn},
  \citenamefont {Lee}, \citenamefont {Kemper}, \citenamefont {Damascelli},
  \citenamefont {Giannetti}, \citenamefont {Turner},\ and\ \citenamefont
  {Coslovich}}]{Wandel2022}%
  \BibitemOpen
  \bibfield  {author} {\bibinfo {author} {\bibfnamefont {S.}~\bibnamefont
  {Wandel}}, \bibinfo {author} {\bibfnamefont {F.}~\bibnamefont {Boschini}},
  \bibinfo {author} {\bibfnamefont {E.~H.}\ \bibnamefont {da~Silva~Neto}},
  \bibinfo {author} {\bibfnamefont {L.}~\bibnamefont {Shen}}, \bibinfo {author}
  {\bibfnamefont {M.~X.}\ \bibnamefont {Na}}, \bibinfo {author} {\bibfnamefont
  {S.}~\bibnamefont {Zohar}}, \bibinfo {author} {\bibfnamefont
  {Y.}~\bibnamefont {Wang}}, \bibinfo {author} {\bibfnamefont {S.~B.}\
  \bibnamefont {Welch}}, \bibinfo {author} {\bibfnamefont {M.~H.}\ \bibnamefont
  {Seaberg}}, \bibinfo {author} {\bibfnamefont {J.~D.}\ \bibnamefont
  {Koralek}}, \bibinfo {author} {\bibfnamefont {G.~L.}\ \bibnamefont
  {Dakovski}}, \bibinfo {author} {\bibfnamefont {W.}~\bibnamefont {Hettel}},
  \bibinfo {author} {\bibfnamefont {M.-F.}\ \bibnamefont {Lin}}, \bibinfo
  {author} {\bibfnamefont {S.~P.}\ \bibnamefont {Moeller}}, \bibinfo {author}
  {\bibfnamefont {W.~F.}\ \bibnamefont {Schlotter}}, \bibinfo {author}
  {\bibfnamefont {A.~H.}\ \bibnamefont {Reid}}, \bibinfo {author}
  {\bibfnamefont {M.~P.}\ \bibnamefont {Minitti}}, \bibinfo {author}
  {\bibfnamefont {T.}~\bibnamefont {Boyle}}, \bibinfo {author} {\bibfnamefont
  {F.}~\bibnamefont {He}}, \bibinfo {author} {\bibfnamefont {R.}~\bibnamefont
  {Sutarto}}, \bibinfo {author} {\bibfnamefont {R.}~\bibnamefont {Liang}},
  \bibinfo {author} {\bibfnamefont {D.}~\bibnamefont {Bonn}}, \bibinfo {author}
  {\bibfnamefont {W.}~\bibnamefont {Hardy}}, \bibinfo {author} {\bibfnamefont
  {R.~A.}\ \bibnamefont {Kaindl}}, \bibinfo {author} {\bibfnamefont {D.~G.}\
  \bibnamefont {Hawthorn}}, \bibinfo {author} {\bibfnamefont {J.-S.}\
  \bibnamefont {Lee}}, \bibinfo {author} {\bibfnamefont {A.~F.}\ \bibnamefont
  {Kemper}}, \bibinfo {author} {\bibfnamefont {A.}~\bibnamefont {Damascelli}},
  \bibinfo {author} {\bibfnamefont {C.}~\bibnamefont {Giannetti}}, \bibinfo
  {author} {\bibfnamefont {J.~J.}\ \bibnamefont {Turner}},\ and\ \bibinfo
  {author} {\bibfnamefont {G.}~\bibnamefont {Coslovich}},\ }\href
  {https://doi.org/10.1126/science.abd7213} {\bibfield  {journal} {\bibinfo
  {journal} {Science}\ }\textbf {\bibinfo {volume} {376}},\ \bibinfo {pages}
  {860} (\bibinfo {year} {2022})}\BibitemShut {NoStop}%
\bibitem [{\citenamefont {Li}\ \emph {et~al.}(2019)\citenamefont {Li},
  \citenamefont {Qiu}, \citenamefont {Zhang}, \citenamefont {Baldini},
  \citenamefont {Lu}, \citenamefont {Rappe},\ and\ \citenamefont
  {Nelson}}]{Li2019}%
  \BibitemOpen
  \bibfield  {author} {\bibinfo {author} {\bibfnamefont {X.}~\bibnamefont
  {Li}}, \bibinfo {author} {\bibfnamefont {T.}~\bibnamefont {Qiu}}, \bibinfo
  {author} {\bibfnamefont {J.}~\bibnamefont {Zhang}}, \bibinfo {author}
  {\bibfnamefont {E.}~\bibnamefont {Baldini}}, \bibinfo {author} {\bibfnamefont
  {J.}~\bibnamefont {Lu}}, \bibinfo {author} {\bibfnamefont {A.~M.}\
  \bibnamefont {Rappe}},\ and\ \bibinfo {author} {\bibfnamefont {K.~A.}\
  \bibnamefont {Nelson}},\ }\href {https://doi.org/10.1126/science.aaw4913}
  {\bibfield  {journal} {\bibinfo  {journal} {Science}\ }\textbf {\bibinfo
  {volume} {364}},\ \bibinfo {pages} {1079} (\bibinfo {year}
  {2019})}\BibitemShut {NoStop}%
\bibitem [{\citenamefont {Nova}\ \emph {et~al.}(2019)\citenamefont {Nova},
  \citenamefont {Disa}, \citenamefont {Fechner},\ and\ \citenamefont
  {Cavalleri}}]{Nova2019}%
  \BibitemOpen
  \bibfield  {author} {\bibinfo {author} {\bibfnamefont {T.~F.}\ \bibnamefont
  {Nova}}, \bibinfo {author} {\bibfnamefont {A.~S.}\ \bibnamefont {Disa}},
  \bibinfo {author} {\bibfnamefont {M.}~\bibnamefont {Fechner}},\ and\ \bibinfo
  {author} {\bibfnamefont {A.}~\bibnamefont {Cavalleri}},\ }\href
  {https://doi.org/10.1126/science.aaw4911} {\bibfield  {journal} {\bibinfo
  {journal} {Science}\ }\textbf {\bibinfo {volume} {364}},\ \bibinfo {pages}
  {1075} (\bibinfo {year} {2019})}\BibitemShut {NoStop}%
\bibitem [{\citenamefont {Stojchevska}\ \emph {et~al.}(2014)\citenamefont
  {Stojchevska}, \citenamefont {Vaskivskyi}, \citenamefont {Mertelj},
  \citenamefont {Kusar}, \citenamefont {Svetin}, \citenamefont {Brazovskii},\
  and\ \citenamefont {Mihailovic}}]{Stojchevska2014}%
  \BibitemOpen
  \bibfield  {author} {\bibinfo {author} {\bibfnamefont {L.}~\bibnamefont
  {Stojchevska}}, \bibinfo {author} {\bibfnamefont {I.}~\bibnamefont
  {Vaskivskyi}}, \bibinfo {author} {\bibfnamefont {T.}~\bibnamefont {Mertelj}},
  \bibinfo {author} {\bibfnamefont {P.}~\bibnamefont {Kusar}}, \bibinfo
  {author} {\bibfnamefont {D.}~\bibnamefont {Svetin}}, \bibinfo {author}
  {\bibfnamefont {S.}~\bibnamefont {Brazovskii}},\ and\ \bibinfo {author}
  {\bibfnamefont {D.}~\bibnamefont {Mihailovic}},\ }\href
  {https://doi.org/10.1126/science.1241591} {\bibfield  {journal} {\bibinfo
  {journal} {Science}\ }\textbf {\bibinfo {volume} {344}},\ \bibinfo {pages}
  {177} (\bibinfo {year} {2014})}\BibitemShut {NoStop}%
\bibitem [{\citenamefont {Morrison}\ \emph {et~al.}(2014)\citenamefont
  {Morrison}, \citenamefont {Chatelain}, \citenamefont {Tiwari}, \citenamefont
  {Hendaoui}, \citenamefont {Bruh$\rm\acute{a}$cs}, \citenamefont {Chaker},\
  and\ \citenamefont {Siwick}}]{Morrison2014}%
  \BibitemOpen
  \bibfield  {author} {\bibinfo {author} {\bibfnamefont {V.~R.}\ \bibnamefont
  {Morrison}}, \bibinfo {author} {\bibfnamefont {R.~P.}\ \bibnamefont
  {Chatelain}}, \bibinfo {author} {\bibfnamefont {K.~L.}\ \bibnamefont
  {Tiwari}}, \bibinfo {author} {\bibfnamefont {A.}~\bibnamefont {Hendaoui}},
  \bibinfo {author} {\bibfnamefont {A.}~\bibnamefont {Bruh$\rm\acute{a}$cs}},
  \bibinfo {author} {\bibfnamefont {M.}~\bibnamefont {Chaker}},\ and\ \bibinfo
  {author} {\bibfnamefont {B.~J.}\ \bibnamefont {Siwick}},\ }\href
  {https://doi.org/10.1126/science.1253779} {\bibfield  {journal} {\bibinfo
  {journal} {Science}\ }\textbf {\bibinfo {volume} {346}},\ \bibinfo {pages}
  {445} (\bibinfo {year} {2014})}\BibitemShut {NoStop}%
\bibitem [{\citenamefont {Kogar}\ \emph {et~al.}(2020)\citenamefont {Kogar},
  \citenamefont {Zong}, \citenamefont {Dolgirev}, \citenamefont {Shen},
  \citenamefont {Straquadine}, \citenamefont {Bie}, \citenamefont {Wang},
  \citenamefont {Rohwer}, \citenamefont {Tung}, \citenamefont {Yang},
  \citenamefont {Li}, \citenamefont {Yang}, \citenamefont {Weathersby},
  \citenamefont {Park}, \citenamefont {Kozina}, \citenamefont {Sie},
  \citenamefont {Wen}, \citenamefont {Jarillo-Herrero}, \citenamefont {Fisher},
  \citenamefont {Wang},\ and\ \citenamefont {Gedik}}]{Kogar2020}%
  \BibitemOpen
  \bibfield  {author} {\bibinfo {author} {\bibfnamefont {A.}~\bibnamefont
  {Kogar}}, \bibinfo {author} {\bibfnamefont {A.}~\bibnamefont {Zong}},
  \bibinfo {author} {\bibfnamefont {P.~E.}\ \bibnamefont {Dolgirev}}, \bibinfo
  {author} {\bibfnamefont {X.}~\bibnamefont {Shen}}, \bibinfo {author}
  {\bibfnamefont {J.}~\bibnamefont {Straquadine}}, \bibinfo {author}
  {\bibfnamefont {Y.-Q.}\ \bibnamefont {Bie}}, \bibinfo {author} {\bibfnamefont
  {X.}~\bibnamefont {Wang}}, \bibinfo {author} {\bibfnamefont {T.}~\bibnamefont
  {Rohwer}}, \bibinfo {author} {\bibfnamefont {I.-C.}\ \bibnamefont {Tung}},
  \bibinfo {author} {\bibfnamefont {Y.}~\bibnamefont {Yang}}, \bibinfo {author}
  {\bibfnamefont {R.}~\bibnamefont {Li}}, \bibinfo {author} {\bibfnamefont
  {J.}~\bibnamefont {Yang}}, \bibinfo {author} {\bibfnamefont {S.}~\bibnamefont
  {Weathersby}}, \bibinfo {author} {\bibfnamefont {S.}~\bibnamefont {Park}},
  \bibinfo {author} {\bibfnamefont {M.~E.}\ \bibnamefont {Kozina}}, \bibinfo
  {author} {\bibfnamefont {E.~J.}\ \bibnamefont {Sie}}, \bibinfo {author}
  {\bibfnamefont {H.}~\bibnamefont {Wen}}, \bibinfo {author} {\bibfnamefont
  {P.}~\bibnamefont {Jarillo-Herrero}}, \bibinfo {author} {\bibfnamefont
  {I.~R.}\ \bibnamefont {Fisher}}, \bibinfo {author} {\bibfnamefont
  {X.}~\bibnamefont {Wang}},\ and\ \bibinfo {author} {\bibfnamefont
  {N.}~\bibnamefont {Gedik}},\ }\href
  {https://doi.org/10.1038/s41567-019-0705-3} {\bibfield  {journal} {\bibinfo
  {journal} {Nature Physics}\ }\textbf {\bibinfo {volume} {16}},\ \bibinfo
  {pages} {159} (\bibinfo {year} {2020})}\BibitemShut {NoStop}%
\bibitem [{\citenamefont {Basov}\ \emph {et~al.}(2011)\citenamefont {Basov},
  \citenamefont {Averitt}, \citenamefont {van~der Marel}, \citenamefont
  {Dressel},\ and\ \citenamefont {Haule}}]{Basov2011}%
  \BibitemOpen
  \bibfield  {author} {\bibinfo {author} {\bibfnamefont {D.~N.}\ \bibnamefont
  {Basov}}, \bibinfo {author} {\bibfnamefont {R.~D.}\ \bibnamefont {Averitt}},
  \bibinfo {author} {\bibfnamefont {D.}~\bibnamefont {van~der Marel}}, \bibinfo
  {author} {\bibfnamefont {M.}~\bibnamefont {Dressel}},\ and\ \bibinfo {author}
  {\bibfnamefont {K.}~\bibnamefont {Haule}},\ }\href
  {https://doi.org/10.1103/RevModPhys.83.471} {\bibfield  {journal} {\bibinfo
  {journal} {Rev. Mod. Phys.}\ }\textbf {\bibinfo {volume} {83}},\ \bibinfo
  {pages} {471} (\bibinfo {year} {2011})}\BibitemShut {NoStop}%
\bibitem [{\citenamefont {Giulietti}\ and\ \citenamefont
  {Ledingham}(2012)}]{Giulietti2012}%
  \BibitemOpen
  \bibfield  {author} {\bibinfo {author} {\bibfnamefont {A.}~\bibnamefont
  {Giulietti}}\ and\ \bibinfo {author} {\bibfnamefont {K.}~\bibnamefont
  {Ledingham}},\ }\href {https://books.google.com/books?id=lE2UtgAACAAJ} {\emph
  {\bibinfo {title} {Progress in Ultrafast Intense Laser Science: Volume V}}},\
  Springer Series in Chemical Physics\ (\bibinfo  {publisher} {Springer Berlin
  Heidelberg},\ \bibinfo {year} {2012})\BibitemShut {NoStop}%
\bibitem [{\citenamefont {Gerber}\ \emph {et~al.}(2017)\citenamefont {Gerber},
  \citenamefont {Yang}, \citenamefont {Zhu}, \citenamefont {Soifer},
  \citenamefont {Sobota}, \citenamefont {Rebec}, \citenamefont {Lee},
  \citenamefont {Jia}, \citenamefont {Moritz}, \citenamefont {Jia},
  \citenamefont {Gauthier}, \citenamefont {Li}, \citenamefont {Leuenberger},
  \citenamefont {Zhang}, \citenamefont {Chaix}, \citenamefont {Li},
  \citenamefont {Jang}, \citenamefont {Lee}, \citenamefont {Yi}, \citenamefont
  {Dakovski}, \citenamefont {Song}, \citenamefont {Glownia}, \citenamefont
  {Nelson}, \citenamefont {Kim}, \citenamefont {Chuang}, \citenamefont
  {Hussain}, \citenamefont {Moore}, \citenamefont {Devereaux}, \citenamefont
  {Lee}, \citenamefont {Kirchmann},\ and\ \citenamefont {Shen}}]{Gerber2017}%
  \BibitemOpen
  \bibfield  {author} {\bibinfo {author} {\bibfnamefont {S.}~\bibnamefont
  {Gerber}}, \bibinfo {author} {\bibfnamefont {S.-L.}\ \bibnamefont {Yang}},
  \bibinfo {author} {\bibfnamefont {D.}~\bibnamefont {Zhu}}, \bibinfo {author}
  {\bibfnamefont {H.}~\bibnamefont {Soifer}}, \bibinfo {author} {\bibfnamefont
  {J.~A.}\ \bibnamefont {Sobota}}, \bibinfo {author} {\bibfnamefont
  {S.}~\bibnamefont {Rebec}}, \bibinfo {author} {\bibfnamefont {J.~J.}\
  \bibnamefont {Lee}}, \bibinfo {author} {\bibfnamefont {T.}~\bibnamefont
  {Jia}}, \bibinfo {author} {\bibfnamefont {B.}~\bibnamefont {Moritz}},
  \bibinfo {author} {\bibfnamefont {C.}~\bibnamefont {Jia}}, \bibinfo {author}
  {\bibfnamefont {A.}~\bibnamefont {Gauthier}}, \bibinfo {author}
  {\bibfnamefont {Y.}~\bibnamefont {Li}}, \bibinfo {author} {\bibfnamefont
  {D.}~\bibnamefont {Leuenberger}}, \bibinfo {author} {\bibfnamefont
  {Y.}~\bibnamefont {Zhang}}, \bibinfo {author} {\bibfnamefont
  {L.}~\bibnamefont {Chaix}}, \bibinfo {author} {\bibfnamefont
  {W.}~\bibnamefont {Li}}, \bibinfo {author} {\bibfnamefont {H.}~\bibnamefont
  {Jang}}, \bibinfo {author} {\bibfnamefont {J.-S.}\ \bibnamefont {Lee}},
  \bibinfo {author} {\bibfnamefont {M.}~\bibnamefont {Yi}}, \bibinfo {author}
  {\bibfnamefont {G.~L.}\ \bibnamefont {Dakovski}}, \bibinfo {author}
  {\bibfnamefont {S.}~\bibnamefont {Song}}, \bibinfo {author} {\bibfnamefont
  {J.~M.}\ \bibnamefont {Glownia}}, \bibinfo {author} {\bibfnamefont
  {S.}~\bibnamefont {Nelson}}, \bibinfo {author} {\bibfnamefont {K.~W.}\
  \bibnamefont {Kim}}, \bibinfo {author} {\bibfnamefont {Y.-D.}\ \bibnamefont
  {Chuang}}, \bibinfo {author} {\bibfnamefont {Z.}~\bibnamefont {Hussain}},
  \bibinfo {author} {\bibfnamefont {R.~G.}\ \bibnamefont {Moore}}, \bibinfo
  {author} {\bibfnamefont {T.~P.}\ \bibnamefont {Devereaux}}, \bibinfo {author}
  {\bibfnamefont {W.-S.}\ \bibnamefont {Lee}}, \bibinfo {author} {\bibfnamefont
  {P.~S.}\ \bibnamefont {Kirchmann}},\ and\ \bibinfo {author} {\bibfnamefont
  {Z.-X.}\ \bibnamefont {Shen}},\ }\href
  {https://doi.org/10.1126/science.aak9946} {\bibfield  {journal} {\bibinfo
  {journal} {Science}\ }\textbf {\bibinfo {volume} {357}},\ \bibinfo {pages}
  {71} (\bibinfo {year} {2017})}\BibitemShut {NoStop}%
\bibitem [{\citenamefont {Yang}\ \emph
  {et~al.}(2019{\natexlab{a}})\citenamefont {Yang}, \citenamefont {Sobota},
  \citenamefont {He}, \citenamefont {Leuenberger}, \citenamefont {Soifer},
  \citenamefont {Eisaki}, \citenamefont {Kirchmann},\ and\ \citenamefont
  {Shen}}]{Yang2019a}%
  \BibitemOpen
  \bibfield  {author} {\bibinfo {author} {\bibfnamefont {S.-L.}\ \bibnamefont
  {Yang}}, \bibinfo {author} {\bibfnamefont {J.~A.}\ \bibnamefont {Sobota}},
  \bibinfo {author} {\bibfnamefont {Y.}~\bibnamefont {He}}, \bibinfo {author}
  {\bibfnamefont {D.}~\bibnamefont {Leuenberger}}, \bibinfo {author}
  {\bibfnamefont {H.}~\bibnamefont {Soifer}}, \bibinfo {author} {\bibfnamefont
  {H.}~\bibnamefont {Eisaki}}, \bibinfo {author} {\bibfnamefont {P.~S.}\
  \bibnamefont {Kirchmann}},\ and\ \bibinfo {author} {\bibfnamefont {Z.-X.}\
  \bibnamefont {Shen}},\ }\href
  {https://doi.org/10.1103/PhysRevLett.122.176403} {\bibfield  {journal}
  {\bibinfo  {journal} {Physical Review Letters}\ }\textbf {\bibinfo {volume}
  {122}},\ \bibinfo {pages} {176403} (\bibinfo {year}
  {2019}{\natexlab{a}})}\BibitemShut {NoStop}%
\bibitem [{\citenamefont {Yang}\ \emph {et~al.}(2022)\citenamefont {Yang},
  \citenamefont {Wang}, \citenamefont {Duan}, \citenamefont {Wo}, \citenamefont
  {Huang}, \citenamefont {Wang}, \citenamefont {Gu}, \citenamefont {Xiang},
  \citenamefont {Qian}, \citenamefont {Zhao},\ and\ \citenamefont
  {Zhang}}]{Yang2022}%
  \BibitemOpen
  \bibfield  {author} {\bibinfo {author} {\bibfnamefont {Y.}~\bibnamefont
  {Yang}}, \bibinfo {author} {\bibfnamefont {Q.}~\bibnamefont {Wang}}, \bibinfo
  {author} {\bibfnamefont {S.}~\bibnamefont {Duan}}, \bibinfo {author}
  {\bibfnamefont {H.}~\bibnamefont {Wo}}, \bibinfo {author} {\bibfnamefont
  {C.}~\bibnamefont {Huang}}, \bibinfo {author} {\bibfnamefont
  {S.}~\bibnamefont {Wang}}, \bibinfo {author} {\bibfnamefont {L.}~\bibnamefont
  {Gu}}, \bibinfo {author} {\bibfnamefont {D.}~\bibnamefont {Xiang}}, \bibinfo
  {author} {\bibfnamefont {D.}~\bibnamefont {Qian}}, \bibinfo {author}
  {\bibfnamefont {J.}~\bibnamefont {Zhao}},\ and\ \bibinfo {author}
  {\bibfnamefont {W.}~\bibnamefont {Zhang}},\ }\href
  {https://doi.org/10.1103/PhysRevLett.128.246401} {\bibfield  {journal}
  {\bibinfo  {journal} {Physical Review Letters}\ }\textbf {\bibinfo {volume}
  {128}},\ \bibinfo {pages} {246401} (\bibinfo {year} {2022})}\BibitemShut
  {NoStop}%
\bibitem [{\citenamefont {Maklar}\ \emph {et~al.}(2021)\citenamefont {Maklar},
  \citenamefont {Windsor}, \citenamefont {Nicholson}, \citenamefont {Puppin},
  \citenamefont {Walmsley}, \citenamefont {Esposito}, \citenamefont {Porer},
  \citenamefont {Rittmann}, \citenamefont {Leuenberger}, \citenamefont {Kubli},
  \citenamefont {Savoini}, \citenamefont {Abreu}, \citenamefont {Johnson},
  \citenamefont {Beaud}, \citenamefont {Ingold}, \citenamefont {Staub},
  \citenamefont {Fisher}, \citenamefont {Ernstorfer}, \citenamefont {Wolf},\
  and\ \citenamefont {Rettig}}]{Maklar2021}%
  \BibitemOpen
  \bibfield  {author} {\bibinfo {author} {\bibfnamefont {J.}~\bibnamefont
  {Maklar}}, \bibinfo {author} {\bibfnamefont {Y.~W.}\ \bibnamefont {Windsor}},
  \bibinfo {author} {\bibfnamefont {C.~W.}\ \bibnamefont {Nicholson}}, \bibinfo
  {author} {\bibfnamefont {M.}~\bibnamefont {Puppin}}, \bibinfo {author}
  {\bibfnamefont {P.}~\bibnamefont {Walmsley}}, \bibinfo {author}
  {\bibfnamefont {V.}~\bibnamefont {Esposito}}, \bibinfo {author}
  {\bibfnamefont {M.}~\bibnamefont {Porer}}, \bibinfo {author} {\bibfnamefont
  {J.}~\bibnamefont {Rittmann}}, \bibinfo {author} {\bibfnamefont
  {D.}~\bibnamefont {Leuenberger}}, \bibinfo {author} {\bibfnamefont
  {M.}~\bibnamefont {Kubli}}, \bibinfo {author} {\bibfnamefont
  {M.}~\bibnamefont {Savoini}}, \bibinfo {author} {\bibfnamefont
  {E.}~\bibnamefont {Abreu}}, \bibinfo {author} {\bibfnamefont {S.~L.}\
  \bibnamefont {Johnson}}, \bibinfo {author} {\bibfnamefont {P.}~\bibnamefont
  {Beaud}}, \bibinfo {author} {\bibfnamefont {G.}~\bibnamefont {Ingold}},
  \bibinfo {author} {\bibfnamefont {U.}~\bibnamefont {Staub}}, \bibinfo
  {author} {\bibfnamefont {I.~R.}\ \bibnamefont {Fisher}}, \bibinfo {author}
  {\bibfnamefont {R.}~\bibnamefont {Ernstorfer}}, \bibinfo {author}
  {\bibfnamefont {M.}~\bibnamefont {Wolf}},\ and\ \bibinfo {author}
  {\bibfnamefont {L.}~\bibnamefont {Rettig}},\ }\href
  {https://doi.org/10.1038/s41467-021-22778-w} {\bibfield  {journal} {\bibinfo
  {journal} {Nature Communications}\ }\textbf {\bibinfo {volume} {12}},\
  \bibinfo {pages} {2499} (\bibinfo {year} {2021})}\BibitemShut {NoStop}%
\bibitem [{\citenamefont {Shi}\ \emph {et~al.}(2019)\citenamefont {Shi},
  \citenamefont {You}, \citenamefont {Zhang}, \citenamefont {Tao},
  \citenamefont {Oppeneer}, \citenamefont {Wu}, \citenamefont {Thomale},
  \citenamefont {Rossnagel}, \citenamefont {Bauer}, \citenamefont {Kapteyn},\
  and\ \citenamefont {Murnane}}]{Shi2019}%
  \BibitemOpen
  \bibfield  {author} {\bibinfo {author} {\bibfnamefont {X.}~\bibnamefont
  {Shi}}, \bibinfo {author} {\bibfnamefont {W.}~\bibnamefont {You}}, \bibinfo
  {author} {\bibfnamefont {Y.}~\bibnamefont {Zhang}}, \bibinfo {author}
  {\bibfnamefont {Z.}~\bibnamefont {Tao}}, \bibinfo {author} {\bibfnamefont
  {P.~M.}\ \bibnamefont {Oppeneer}}, \bibinfo {author} {\bibfnamefont
  {X.}~\bibnamefont {Wu}}, \bibinfo {author} {\bibfnamefont {R.}~\bibnamefont
  {Thomale}}, \bibinfo {author} {\bibfnamefont {K.}~\bibnamefont {Rossnagel}},
  \bibinfo {author} {\bibfnamefont {M.}~\bibnamefont {Bauer}}, \bibinfo
  {author} {\bibfnamefont {H.}~\bibnamefont {Kapteyn}},\ and\ \bibinfo {author}
  {\bibfnamefont {M.}~\bibnamefont {Murnane}},\ }\href
  {https://doi.org/10.1126/sciadv.aav4449} {\bibfield  {journal} {\bibinfo
  {journal} {Science Advances}\ }\textbf {\bibinfo {volume} {5}},\ \bibinfo
  {pages} {eaav4449} (\bibinfo {year} {2019})}\BibitemShut {NoStop}%
\bibitem [{\citenamefont {Duan}\ \emph {et~al.}(2021)\citenamefont {Duan},
  \citenamefont {Cheng}, \citenamefont {Xia}, \citenamefont {Yang},
  \citenamefont {Xu}, \citenamefont {Qi}, \citenamefont {Huang}, \citenamefont
  {Tang}, \citenamefont {Guo}, \citenamefont {Luo}, \citenamefont {Qian},
  \citenamefont {Xiang}, \citenamefont {Zhang},\ and\ \citenamefont
  {Zhang}}]{Duan2021}%
  \BibitemOpen
  \bibfield  {author} {\bibinfo {author} {\bibfnamefont {S.}~\bibnamefont
  {Duan}}, \bibinfo {author} {\bibfnamefont {Y.}~\bibnamefont {Cheng}},
  \bibinfo {author} {\bibfnamefont {W.}~\bibnamefont {Xia}}, \bibinfo {author}
  {\bibfnamefont {Y.}~\bibnamefont {Yang}}, \bibinfo {author} {\bibfnamefont
  {C.}~\bibnamefont {Xu}}, \bibinfo {author} {\bibfnamefont {F.}~\bibnamefont
  {Qi}}, \bibinfo {author} {\bibfnamefont {C.}~\bibnamefont {Huang}}, \bibinfo
  {author} {\bibfnamefont {T.}~\bibnamefont {Tang}}, \bibinfo {author}
  {\bibfnamefont {Y.}~\bibnamefont {Guo}}, \bibinfo {author} {\bibfnamefont
  {W.}~\bibnamefont {Luo}}, \bibinfo {author} {\bibfnamefont {D.}~\bibnamefont
  {Qian}}, \bibinfo {author} {\bibfnamefont {D.}~\bibnamefont {Xiang}},
  \bibinfo {author} {\bibfnamefont {J.}~\bibnamefont {Zhang}},\ and\ \bibinfo
  {author} {\bibfnamefont {W.}~\bibnamefont {Zhang}},\ }\href
  {https://doi.org/10.1038/s41586-021-03643-8} {\bibfield  {journal} {\bibinfo
  {journal} {Nature}\ }\textbf {\bibinfo {volume} {595}},\ \bibinfo {pages}
  {239} (\bibinfo {year} {2021})}\BibitemShut {NoStop}%
\bibitem [{\citenamefont {Sobota}\ \emph {et~al.}(2014)\citenamefont {Sobota},
  \citenamefont {Yang}, \citenamefont {Leuenberger}, \citenamefont {Kemper},
  \citenamefont {Analytis}, \citenamefont {Fisher}, \citenamefont {Kirchmann},
  \citenamefont {Devereaux},\ and\ \citenamefont {Shen}}]{Sobota2014}%
  \BibitemOpen
  \bibfield  {author} {\bibinfo {author} {\bibfnamefont {J.~A.}\ \bibnamefont
  {Sobota}}, \bibinfo {author} {\bibfnamefont {S.-L.}\ \bibnamefont {Yang}},
  \bibinfo {author} {\bibfnamefont {D.}~\bibnamefont {Leuenberger}}, \bibinfo
  {author} {\bibfnamefont {A.~F.}\ \bibnamefont {Kemper}}, \bibinfo {author}
  {\bibfnamefont {J.~G.}\ \bibnamefont {Analytis}}, \bibinfo {author}
  {\bibfnamefont {I.~R.}\ \bibnamefont {Fisher}}, \bibinfo {author}
  {\bibfnamefont {P.~S.}\ \bibnamefont {Kirchmann}}, \bibinfo {author}
  {\bibfnamefont {T.~P.}\ \bibnamefont {Devereaux}},\ and\ \bibinfo {author}
  {\bibfnamefont {Z.-X.}\ \bibnamefont {Shen}},\ }\href
  {https://doi.org/10.1103/PhysRevLett.113.157401} {\bibfield  {journal}
  {\bibinfo  {journal} {Physical Review Letters}\ }\textbf {\bibinfo {volume}
  {113}},\ \bibinfo {pages} {157401} (\bibinfo {year} {2014})}\BibitemShut
  {NoStop}%
\bibitem [{\citenamefont {Bretscher}\ \emph {et~al.}(2021)\citenamefont
  {Bretscher}, \citenamefont {Andrich}, \citenamefont {Telang}, \citenamefont
  {Singh}, \citenamefont {Harnagea}, \citenamefont {Sood},\ and\ \citenamefont
  {Rao}}]{Bretscher2021}%
  \BibitemOpen
  \bibfield  {author} {\bibinfo {author} {\bibfnamefont {H.~M.}\ \bibnamefont
  {Bretscher}}, \bibinfo {author} {\bibfnamefont {P.}~\bibnamefont {Andrich}},
  \bibinfo {author} {\bibfnamefont {P.}~\bibnamefont {Telang}}, \bibinfo
  {author} {\bibfnamefont {A.}~\bibnamefont {Singh}}, \bibinfo {author}
  {\bibfnamefont {L.}~\bibnamefont {Harnagea}}, \bibinfo {author}
  {\bibfnamefont {A.~K.}\ \bibnamefont {Sood}},\ and\ \bibinfo {author}
  {\bibfnamefont {A.}~\bibnamefont {Rao}},\ }\href
  {https://doi.org/10.1038/s41467-021-21929-3} {\bibfield  {journal} {\bibinfo
  {journal} {Nature Communications}\ }\textbf {\bibinfo {volume} {12}},\
  \bibinfo {pages} {1699} (\bibinfo {year} {2021})}\BibitemShut {NoStop}%
\bibitem [{\citenamefont {Tang}\ \emph {et~al.}(2020)\citenamefont {Tang},
  \citenamefont {Wang}, \citenamefont {Duan}, \citenamefont {Yang},
  \citenamefont {Huang}, \citenamefont {Guo}, \citenamefont {Qian},\ and\
  \citenamefont {Zhang}}]{Tang2020}%
  \BibitemOpen
  \bibfield  {author} {\bibinfo {author} {\bibfnamefont {T.}~\bibnamefont
  {Tang}}, \bibinfo {author} {\bibfnamefont {H.}~\bibnamefont {Wang}}, \bibinfo
  {author} {\bibfnamefont {S.}~\bibnamefont {Duan}}, \bibinfo {author}
  {\bibfnamefont {Y.}~\bibnamefont {Yang}}, \bibinfo {author} {\bibfnamefont
  {C.}~\bibnamefont {Huang}}, \bibinfo {author} {\bibfnamefont
  {Y.}~\bibnamefont {Guo}}, \bibinfo {author} {\bibfnamefont {D.}~\bibnamefont
  {Qian}},\ and\ \bibinfo {author} {\bibfnamefont {W.}~\bibnamefont {Zhang}},\
  }\href {https://doi.org/10.1103/PhysRevB.101.235148} {\bibfield  {journal}
  {\bibinfo  {journal} {Physical Review B}\ }\textbf {\bibinfo {volume}
  {101}},\ \bibinfo {pages} {235148} (\bibinfo {year} {2020})}\BibitemShut
  {NoStop}%
\bibitem [{\citenamefont {Di~Salvo}\ \emph {et~al.}(1976)\citenamefont
  {Di~Salvo}, \citenamefont {Moncton},\ and\ \citenamefont
  {Waszczak}}]{DiSalvo1976}%
  \BibitemOpen
  \bibfield  {author} {\bibinfo {author} {\bibfnamefont {F.~J.}\ \bibnamefont
  {Di~Salvo}}, \bibinfo {author} {\bibfnamefont {D.~E.}\ \bibnamefont
  {Moncton}},\ and\ \bibinfo {author} {\bibfnamefont {J.~V.}\ \bibnamefont
  {Waszczak}},\ }\href {https://doi.org/10.1103/PhysRevB.14.4321} {\bibfield
  {journal} {\bibinfo  {journal} {Physical Review B}\ }\textbf {\bibinfo
  {volume} {14}},\ \bibinfo {pages} {4321} (\bibinfo {year}
  {1976})}\BibitemShut {NoStop}%
\bibitem [{\citenamefont {Kogar}\ \emph {et~al.}(2017)\citenamefont {Kogar},
  \citenamefont {Rak}, \citenamefont {Vig}, \citenamefont {Husain},
  \citenamefont {Flicker}, \citenamefont {Joe}, \citenamefont {Venema},
  \citenamefont {MacDougall}, \citenamefont {Chiang}, \citenamefont {Fradkin},
  \citenamefont {van Wezel},\ and\ \citenamefont {Abbamonte}}]{Kogar2017}%
  \BibitemOpen
  \bibfield  {author} {\bibinfo {author} {\bibfnamefont {A.}~\bibnamefont
  {Kogar}}, \bibinfo {author} {\bibfnamefont {M.~S.}\ \bibnamefont {Rak}},
  \bibinfo {author} {\bibfnamefont {S.}~\bibnamefont {Vig}}, \bibinfo {author}
  {\bibfnamefont {A.~A.}\ \bibnamefont {Husain}}, \bibinfo {author}
  {\bibfnamefont {F.}~\bibnamefont {Flicker}}, \bibinfo {author} {\bibfnamefont
  {Y.~I.}\ \bibnamefont {Joe}}, \bibinfo {author} {\bibfnamefont
  {L.}~\bibnamefont {Venema}}, \bibinfo {author} {\bibfnamefont {G.~J.}\
  \bibnamefont {MacDougall}}, \bibinfo {author} {\bibfnamefont {T.~C.}\
  \bibnamefont {Chiang}}, \bibinfo {author} {\bibfnamefont {E.}~\bibnamefont
  {Fradkin}}, \bibinfo {author} {\bibfnamefont {J.}~\bibnamefont {van Wezel}},\
  and\ \bibinfo {author} {\bibfnamefont {P.}~\bibnamefont {Abbamonte}},\ }\href
  {https://doi.org/10.1126/science.aam6432} {\bibfield  {journal} {\bibinfo
  {journal} {Science}\ }\textbf {\bibinfo {volume} {358}},\ \bibinfo {pages}
  {1314} (\bibinfo {year} {2017})}\BibitemShut {NoStop}%
\bibitem [{\citenamefont {Porer}\ \emph {et~al.}(2014)\citenamefont {Porer},
  \citenamefont {Leierseder}, \citenamefont {M$\rm\acute{e}$nard},
  \citenamefont {Dachraoui}, \citenamefont {Mouchliadis}, \citenamefont
  {Perakis}, \citenamefont {Heinzmann}, \citenamefont {Demsar}, \citenamefont
  {Rossnagel},\ and\ \citenamefont {Huber}}]{Porer2014}%
  \BibitemOpen
  \bibfield  {author} {\bibinfo {author} {\bibfnamefont {M.}~\bibnamefont
  {Porer}}, \bibinfo {author} {\bibfnamefont {U.}~\bibnamefont {Leierseder}},
  \bibinfo {author} {\bibfnamefont {J.-M.}\ \bibnamefont
  {M$\rm\acute{e}$nard}}, \bibinfo {author} {\bibfnamefont {H.}~\bibnamefont
  {Dachraoui}}, \bibinfo {author} {\bibfnamefont {L.}~\bibnamefont
  {Mouchliadis}}, \bibinfo {author} {\bibfnamefont {I.~E.}\ \bibnamefont
  {Perakis}}, \bibinfo {author} {\bibfnamefont {U.}~\bibnamefont {Heinzmann}},
  \bibinfo {author} {\bibfnamefont {J.}~\bibnamefont {Demsar}}, \bibinfo
  {author} {\bibfnamefont {K.}~\bibnamefont {Rossnagel}},\ and\ \bibinfo
  {author} {\bibfnamefont {R.}~\bibnamefont {Huber}},\ }\href
  {https://doi.org/10.1038/nmat4042} {\bibfield  {journal} {\bibinfo  {journal}
  {Nature Materials}\ }\textbf {\bibinfo {volume} {13}},\ \bibinfo {pages}
  {857} (\bibinfo {year} {2014})}\BibitemShut {NoStop}%
\bibitem [{\citenamefont {Cheng}\ \emph {et~al.}(2022)\citenamefont {Cheng},
  \citenamefont {Zong}, \citenamefont {Li}, \citenamefont {Xia}, \citenamefont
  {Duan}, \citenamefont {Zhao}, \citenamefont {Li}, \citenamefont {Qi},
  \citenamefont {Wu}, \citenamefont {Zhao}, \citenamefont {Zhu}, \citenamefont
  {Zou}, \citenamefont {Jiang}, \citenamefont {Guo}, \citenamefont {Yang},
  \citenamefont {Qian}, \citenamefont {Zhang}, \citenamefont {Kogar},
  \citenamefont {Zuerch}, \citenamefont {Xiang},\ and\ \citenamefont
  {Zhang}}]{Cheng2022}%
  \BibitemOpen
  \bibfield  {author} {\bibinfo {author} {\bibfnamefont {Y.}~\bibnamefont
  {Cheng}}, \bibinfo {author} {\bibfnamefont {A.}~\bibnamefont {Zong}},
  \bibinfo {author} {\bibfnamefont {J.}~\bibnamefont {Li}}, \bibinfo {author}
  {\bibfnamefont {W.}~\bibnamefont {Xia}}, \bibinfo {author} {\bibfnamefont
  {S.}~\bibnamefont {Duan}}, \bibinfo {author} {\bibfnamefont {W.}~\bibnamefont
  {Zhao}}, \bibinfo {author} {\bibfnamefont {Y.}~\bibnamefont {Li}}, \bibinfo
  {author} {\bibfnamefont {F.}~\bibnamefont {Qi}}, \bibinfo {author}
  {\bibfnamefont {J.}~\bibnamefont {Wu}}, \bibinfo {author} {\bibfnamefont
  {L.}~\bibnamefont {Zhao}}, \bibinfo {author} {\bibfnamefont {P.}~\bibnamefont
  {Zhu}}, \bibinfo {author} {\bibfnamefont {X.}~\bibnamefont {Zou}}, \bibinfo
  {author} {\bibfnamefont {T.}~\bibnamefont {Jiang}}, \bibinfo {author}
  {\bibfnamefont {Y.}~\bibnamefont {Guo}}, \bibinfo {author} {\bibfnamefont
  {L.}~\bibnamefont {Yang}}, \bibinfo {author} {\bibfnamefont {D.}~\bibnamefont
  {Qian}}, \bibinfo {author} {\bibfnamefont {W.}~\bibnamefont {Zhang}},
  \bibinfo {author} {\bibfnamefont {A.}~\bibnamefont {Kogar}}, \bibinfo
  {author} {\bibfnamefont {M.~W.}\ \bibnamefont {Zuerch}}, \bibinfo {author}
  {\bibfnamefont {D.}~\bibnamefont {Xiang}},\ and\ \bibinfo {author}
  {\bibfnamefont {J.}~\bibnamefont {Zhang}},\ }\href
  {https://doi.org/10.1038/s41467-022-28309-5} {\bibfield  {journal} {\bibinfo
  {journal} {Nature Communications}\ }\textbf {\bibinfo {volume} {13}},\
  \bibinfo {pages} {963} (\bibinfo {year} {2022})}\BibitemShut {NoStop}%
\bibitem [{\citenamefont {Cercellier}\ \emph {et~al.}(2007)\citenamefont
  {Cercellier}, \citenamefont {Monney}, \citenamefont {Clerc}, \citenamefont
  {Battaglia}, \citenamefont {Despont}, \citenamefont {Garnier}, \citenamefont
  {Beck}, \citenamefont {Aebi}, \citenamefont {Patthey}, \citenamefont
  {Berger},\ and\ \citenamefont {Forr$\rm\acute{o}$}}]{Cercellier2007}%
  \BibitemOpen
  \bibfield  {author} {\bibinfo {author} {\bibfnamefont {H.}~\bibnamefont
  {Cercellier}}, \bibinfo {author} {\bibfnamefont {C.}~\bibnamefont {Monney}},
  \bibinfo {author} {\bibfnamefont {F.}~\bibnamefont {Clerc}}, \bibinfo
  {author} {\bibfnamefont {C.}~\bibnamefont {Battaglia}}, \bibinfo {author}
  {\bibfnamefont {L.}~\bibnamefont {Despont}}, \bibinfo {author} {\bibfnamefont
  {M.~G.}\ \bibnamefont {Garnier}}, \bibinfo {author} {\bibfnamefont
  {H.}~\bibnamefont {Beck}}, \bibinfo {author} {\bibfnamefont {P.}~\bibnamefont
  {Aebi}}, \bibinfo {author} {\bibfnamefont {L.}~\bibnamefont {Patthey}},
  \bibinfo {author} {\bibfnamefont {H.}~\bibnamefont {Berger}},\ and\ \bibinfo
  {author} {\bibfnamefont {L.}~\bibnamefont {Forr$\rm\acute{o}$}},\ }\href
  {https://doi.org/10.1103/PhysRevLett.99.146403} {\bibfield  {journal}
  {\bibinfo  {journal} {Physical Review Letters}\ }\textbf {\bibinfo {volume}
  {99}},\ \bibinfo {pages} {146403} (\bibinfo {year} {2007})}\BibitemShut
  {NoStop}%
\bibitem [{\citenamefont {Hedayat}\ \emph {et~al.}(2019)\citenamefont
  {Hedayat}, \citenamefont {Sayers}, \citenamefont {Bugini}, \citenamefont
  {Dallera}, \citenamefont {Wolverson}, \citenamefont {Batten}, \citenamefont
  {Karbassi}, \citenamefont {Friedemann}, \citenamefont {Cerullo},
  \citenamefont {van Wezel}, \citenamefont {Clark}, \citenamefont {Carpene},\
  and\ \citenamefont {Da~Como}}]{Hedayat2019}%
  \BibitemOpen
  \bibfield  {author} {\bibinfo {author} {\bibfnamefont {H.}~\bibnamefont
  {Hedayat}}, \bibinfo {author} {\bibfnamefont {C.~J.}\ \bibnamefont {Sayers}},
  \bibinfo {author} {\bibfnamefont {D.}~\bibnamefont {Bugini}}, \bibinfo
  {author} {\bibfnamefont {C.}~\bibnamefont {Dallera}}, \bibinfo {author}
  {\bibfnamefont {D.}~\bibnamefont {Wolverson}}, \bibinfo {author}
  {\bibfnamefont {T.}~\bibnamefont {Batten}}, \bibinfo {author} {\bibfnamefont
  {S.}~\bibnamefont {Karbassi}}, \bibinfo {author} {\bibfnamefont
  {S.}~\bibnamefont {Friedemann}}, \bibinfo {author} {\bibfnamefont
  {G.}~\bibnamefont {Cerullo}}, \bibinfo {author} {\bibfnamefont
  {J.}~\bibnamefont {van Wezel}}, \bibinfo {author} {\bibfnamefont {S.~R.}\
  \bibnamefont {Clark}}, \bibinfo {author} {\bibfnamefont {E.}~\bibnamefont
  {Carpene}},\ and\ \bibinfo {author} {\bibfnamefont {E.}~\bibnamefont
  {Da~Como}},\ }\href {https://doi.org/10.1103/PhysRevResearch.1.023029}
  {\bibfield  {journal} {\bibinfo  {journal} {Physical Review Research}\
  }\textbf {\bibinfo {volume} {1}},\ \bibinfo {pages} {023029} (\bibinfo {year}
  {2019})}\BibitemShut {NoStop}%
\bibitem [{\citenamefont {Morosan}\ \emph {et~al.}(2006)\citenamefont
  {Morosan}, \citenamefont {Zandbergen}, \citenamefont {Dennis}, \citenamefont
  {Bos}, \citenamefont {Onose}, \citenamefont {Klimczuk}, \citenamefont
  {Ramirez}, \citenamefont {Ong},\ and\ \citenamefont {Cava}}]{Morosan2006}%
  \BibitemOpen
  \bibfield  {author} {\bibinfo {author} {\bibfnamefont {E.}~\bibnamefont
  {Morosan}}, \bibinfo {author} {\bibfnamefont {H.~W.}\ \bibnamefont
  {Zandbergen}}, \bibinfo {author} {\bibfnamefont {B.~S.}\ \bibnamefont
  {Dennis}}, \bibinfo {author} {\bibfnamefont {J.~W.~G.}\ \bibnamefont {Bos}},
  \bibinfo {author} {\bibfnamefont {Y.}~\bibnamefont {Onose}}, \bibinfo
  {author} {\bibfnamefont {T.}~\bibnamefont {Klimczuk}}, \bibinfo {author}
  {\bibfnamefont {A.~P.}\ \bibnamefont {Ramirez}}, \bibinfo {author}
  {\bibfnamefont {N.~P.}\ \bibnamefont {Ong}},\ and\ \bibinfo {author}
  {\bibfnamefont {R.~J.}\ \bibnamefont {Cava}},\ }\href
  {https://doi.org/10.1038/nphys360} {\bibfield  {journal} {\bibinfo  {journal}
  {Nature Physics}\ }\textbf {\bibinfo {volume} {2}},\ \bibinfo {pages} {544}
  (\bibinfo {year} {2006})}\BibitemShut {NoStop}%
\bibitem [{\citenamefont {Kusmartseva}\ \emph {et~al.}(2009)\citenamefont
  {Kusmartseva}, \citenamefont {Sipos}, \citenamefont {Berger}, \citenamefont
  {Forr$\rm\acute{o}~$},\ and\ \citenamefont
  {Tuti$\rm\check{s}~$}}]{Kusmartseva2009}%
  \BibitemOpen
  \bibfield  {author} {\bibinfo {author} {\bibfnamefont {A.~F.}\ \bibnamefont
  {Kusmartseva}}, \bibinfo {author} {\bibfnamefont {B.}~\bibnamefont {Sipos}},
  \bibinfo {author} {\bibfnamefont {H.}~\bibnamefont {Berger}}, \bibinfo
  {author} {\bibfnamefont {L.}~\bibnamefont {Forr$\rm\acute{o}~$}},\ and\
  \bibinfo {author} {\bibfnamefont {E.}~\bibnamefont {Tutis}},\
  }\href {https://doi.org/10.1103/PhysRevLett.103.236401} {\bibfield  {journal}
  {\bibinfo  {journal} {Physical Review Letters}\ }\textbf {\bibinfo {volume}
  {103}},\ \bibinfo {pages} {236401} (\bibinfo {year} {2009})}\BibitemShut
  {NoStop}%
\bibitem [{\citenamefont {Joe}\ \emph {et~al.}(2014)\citenamefont {Joe},
  \citenamefont {Chen}, \citenamefont {Ghaemi}, \citenamefont {Finkelstein},
  \citenamefont {de~la Pe$\rm\tilde{n}$a}, \citenamefont {Gan}, \citenamefont
  {Lee}, \citenamefont {Yuan}, \citenamefont {Geck}, \citenamefont
  {MacDougall}, \citenamefont {Chiang}, \citenamefont {Cooper}, \citenamefont
  {Fradkin},\ and\ \citenamefont {Abbamonte}}]{Joe2014}%
  \BibitemOpen
  \bibfield  {author} {\bibinfo {author} {\bibfnamefont {Y.~I.}\ \bibnamefont
  {Joe}}, \bibinfo {author} {\bibfnamefont {X.~M.}\ \bibnamefont {Chen}},
  \bibinfo {author} {\bibfnamefont {P.}~\bibnamefont {Ghaemi}}, \bibinfo
  {author} {\bibfnamefont {K.~D.}\ \bibnamefont {Finkelstein}}, \bibinfo
  {author} {\bibfnamefont {G.~A.}\ \bibnamefont {de~la Pe$\rm\tilde{n}$a}},
  \bibinfo {author} {\bibfnamefont {Y.}~\bibnamefont {Gan}}, \bibinfo {author}
  {\bibfnamefont {J.~C.~T.}\ \bibnamefont {Lee}}, \bibinfo {author}
  {\bibfnamefont {S.}~\bibnamefont {Yuan}}, \bibinfo {author} {\bibfnamefont
  {J.}~\bibnamefont {Geck}}, \bibinfo {author} {\bibfnamefont {G.~J.}\
  \bibnamefont {MacDougall}}, \bibinfo {author} {\bibfnamefont {T.~C.}\
  \bibnamefont {Chiang}}, \bibinfo {author} {\bibfnamefont {S.~L.}\
  \bibnamefont {Cooper}}, \bibinfo {author} {\bibfnamefont {E.}~\bibnamefont
  {Fradkin}},\ and\ \bibinfo {author} {\bibfnamefont {P.}~\bibnamefont
  {Abbamonte}},\ }\href {https://doi.org/10.1038/nphys2935} {\bibfield
  {journal} {\bibinfo  {journal} {Nature Physics}\ }\textbf {\bibinfo {volume}
  {10}},\ \bibinfo {pages} {421} (\bibinfo {year} {2014})}\BibitemShut
  {NoStop}%
\bibitem [{\citenamefont {Li}\ \emph {et~al.}(2016)\citenamefont {Li},
  \citenamefont {O'Farrell}, \citenamefont {Loh}, \citenamefont {Eda},
  \citenamefont {$\rm\ddot{O}$zyilmaz},\ and\ \citenamefont
  {Castro~Neto}}]{Li2016}%
  \BibitemOpen
  \bibfield  {author} {\bibinfo {author} {\bibfnamefont {L.~J.}\ \bibnamefont
  {Li}}, \bibinfo {author} {\bibfnamefont {E.~C.~T.}\ \bibnamefont
  {O'Farrell}}, \bibinfo {author} {\bibfnamefont {K.~P.}\ \bibnamefont {Loh}},
  \bibinfo {author} {\bibfnamefont {G.}~\bibnamefont {Eda}}, \bibinfo {author}
  {\bibfnamefont {B.}~\bibnamefont {$\rm\ddot{O}$zyilmaz}},\ and\ \bibinfo
  {author} {\bibfnamefont {A.~H.}\ \bibnamefont {Castro~Neto}},\ }\href
  {https://doi.org/10.1038/nature16175} {\bibfield  {journal} {\bibinfo
  {journal} {Nature}\ }\textbf {\bibinfo {volume} {529}},\ \bibinfo {pages}
  {185} (\bibinfo {year} {2016})}\BibitemShut {NoStop}%
\bibitem [{\citenamefont {Xu}\ \emph {et~al.}(2020)\citenamefont {Xu},
  \citenamefont {Ma}, \citenamefont {Gao}, \citenamefont {Kogar}, \citenamefont
  {Zong}, \citenamefont {Mier~Valdivia}, \citenamefont {Dinh}, \citenamefont
  {Huang}, \citenamefont {Singh}, \citenamefont {Hsu}, \citenamefont {Chang},
  \citenamefont {Ruff}, \citenamefont {Watanabe}, \citenamefont {Taniguchi},
  \citenamefont {Lin}, \citenamefont {Karapetrov}, \citenamefont {Xiao},
  \citenamefont {Jarillo-Herrero},\ and\ \citenamefont {Gedik}}]{Xu2020}%
  \BibitemOpen
  \bibfield  {author} {\bibinfo {author} {\bibfnamefont {S.-Y.}\ \bibnamefont
  {Xu}}, \bibinfo {author} {\bibfnamefont {Q.}~\bibnamefont {Ma}}, \bibinfo
  {author} {\bibfnamefont {Y.}~\bibnamefont {Gao}}, \bibinfo {author}
  {\bibfnamefont {A.}~\bibnamefont {Kogar}}, \bibinfo {author} {\bibfnamefont
  {A.}~\bibnamefont {Zong}}, \bibinfo {author} {\bibfnamefont {A.~M.}\
  \bibnamefont {Mier~Valdivia}}, \bibinfo {author} {\bibfnamefont {T.~H.}\
  \bibnamefont {Dinh}}, \bibinfo {author} {\bibfnamefont {S.-M.}\ \bibnamefont
  {Huang}}, \bibinfo {author} {\bibfnamefont {B.}~\bibnamefont {Singh}},
  \bibinfo {author} {\bibfnamefont {C.-H.}\ \bibnamefont {Hsu}}, \bibinfo
  {author} {\bibfnamefont {T.-R.}\ \bibnamefont {Chang}}, \bibinfo {author}
  {\bibfnamefont {J.~P.~C.}\ \bibnamefont {Ruff}}, \bibinfo {author}
  {\bibfnamefont {K.}~\bibnamefont {Watanabe}}, \bibinfo {author}
  {\bibfnamefont {T.}~\bibnamefont {Taniguchi}}, \bibinfo {author}
  {\bibfnamefont {H.}~\bibnamefont {Lin}}, \bibinfo {author} {\bibfnamefont
  {G.}~\bibnamefont {Karapetrov}}, \bibinfo {author} {\bibfnamefont
  {D.}~\bibnamefont {Xiao}}, \bibinfo {author} {\bibfnamefont {P.}~\bibnamefont
  {Jarillo-Herrero}},\ and\ \bibinfo {author} {\bibfnamefont {N.}~\bibnamefont
  {Gedik}},\ }\href {https://doi.org/10.1038/s41586-020-2011-8} {\bibfield
  {journal} {\bibinfo  {journal} {Nature}\ }\textbf {\bibinfo {volume} {578}},\
  \bibinfo {pages} {545} (\bibinfo {year} {2020})}\BibitemShut {NoStop}%
\bibitem [{\citenamefont {Wickramaratne}\ \emph {et~al.}(2022)\citenamefont
  {Wickramaratne}, \citenamefont {Subedi}, \citenamefont {Torchinsky},
  \citenamefont {Karapetrov},\ and\ \citenamefont {Mazin}}]{Wickramaratne2022}%
  \BibitemOpen
  \bibfield  {author} {\bibinfo {author} {\bibfnamefont {D.}~\bibnamefont
  {Wickramaratne}}, \bibinfo {author} {\bibfnamefont {S.}~\bibnamefont
  {Subedi}}, \bibinfo {author} {\bibfnamefont {D.~H.}\ \bibnamefont
  {Torchinsky}}, \bibinfo {author} {\bibfnamefont {G.}~\bibnamefont
  {Karapetrov}},\ and\ \bibinfo {author} {\bibfnamefont {I.~I.}\ \bibnamefont
  {Mazin}},\ }\href {https://doi.org/10.1103/PhysRevB.105.054102} {\bibfield
  {journal} {\bibinfo  {journal} {Physical Review B}\ }\textbf {\bibinfo
  {volume} {105}},\ \bibinfo {pages} {054102} (\bibinfo {year}
  {2022})}\BibitemShut {NoStop}%
\bibitem [{\citenamefont {Holy}\ \emph {et~al.}(1977)\citenamefont {Holy},
  \citenamefont {Woo}, \citenamefont {Klein},\ and\ \citenamefont
  {Brown}}]{Holy1977}%
  \BibitemOpen
  \bibfield  {author} {\bibinfo {author} {\bibfnamefont {J.~A.}\ \bibnamefont
  {Holy}}, \bibinfo {author} {\bibfnamefont {K.~C.}\ \bibnamefont {Woo}},
  \bibinfo {author} {\bibfnamefont {M.~V.}\ \bibnamefont {Klein}},\ and\
  \bibinfo {author} {\bibfnamefont {F.~C.}\ \bibnamefont {Brown}},\ }\href
  {https://doi.org/10.1103/PhysRevB.16.3628} {\bibfield  {journal} {\bibinfo
  {journal} {Physical Review B}\ }\textbf {\bibinfo {volume} {16}},\ \bibinfo
  {pages} {3628} (\bibinfo {year} {1977})}\BibitemShut {NoStop}%
\bibitem [{\citenamefont {Snow}\ \emph {et~al.}(2003)\citenamefont {Snow},
  \citenamefont {Karpus}, \citenamefont {Cooper}, \citenamefont {Kidd},\ and\
  \citenamefont {Chiang}}]{Snow2003}%
  \BibitemOpen
  \bibfield  {author} {\bibinfo {author} {\bibfnamefont {C.~S.}\ \bibnamefont
  {Snow}}, \bibinfo {author} {\bibfnamefont {J.~F.}\ \bibnamefont {Karpus}},
  \bibinfo {author} {\bibfnamefont {S.~L.}\ \bibnamefont {Cooper}}, \bibinfo
  {author} {\bibfnamefont {T.~E.}\ \bibnamefont {Kidd}},\ and\ \bibinfo
  {author} {\bibfnamefont {T.-C.}\ \bibnamefont {Chiang}},\ }\href
  {https://doi.org/10.1103/PhysRevLett.91.136402} {\bibfield  {journal}
  {\bibinfo  {journal} {Physical Review Letters}\ }\textbf {\bibinfo {volume}
  {91}},\ \bibinfo {pages} {136402} (\bibinfo {year} {2003})}\BibitemShut
  {NoStop}%
\bibitem [{\citenamefont {Yang}\ \emph
  {et~al.}(2019{\natexlab{b}})\citenamefont {Yang}, \citenamefont {Tang},
  \citenamefont {Duan}, \citenamefont {Zhou}, \citenamefont {Hao},\ and\
  \citenamefont {Zhang}}]{Yang2019}%
  \BibitemOpen
  \bibfield  {author} {\bibinfo {author} {\bibfnamefont {Y.}~\bibnamefont
  {Yang}}, \bibinfo {author} {\bibfnamefont {T.}~\bibnamefont {Tang}}, \bibinfo
  {author} {\bibfnamefont {S.}~\bibnamefont {Duan}}, \bibinfo {author}
  {\bibfnamefont {C.}~\bibnamefont {Zhou}}, \bibinfo {author} {\bibfnamefont
  {D.}~\bibnamefont {Hao}},\ and\ \bibinfo {author} {\bibfnamefont
  {W.}~\bibnamefont {Zhang}},\ }\href {https://doi.org/10.1063/1.5090439}
  {\bibfield  {journal} {\bibinfo  {journal} {Review of Scientific
  Instruments}\ }\textbf {\bibinfo {volume} {90}},\ \bibinfo {pages} {063905}
  (\bibinfo {year} {2019}{\natexlab{b}})}\BibitemShut {NoStop}%
\bibitem [{\citenamefont {Huang}\ \emph {et~al.}(2022)\citenamefont {Huang},
  \citenamefont {Duan},\ and\ \citenamefont {Zhang}}]{Huang2022}%
  \BibitemOpen
  \bibfield  {author} {\bibinfo {author} {\bibfnamefont {C.}~\bibnamefont
  {Huang}}, \bibinfo {author} {\bibfnamefont {S.}~\bibnamefont {Duan}},\ and\
  \bibinfo {author} {\bibfnamefont {W.}~\bibnamefont {Zhang}},\ }\href
  {https://doi.org/10.1007/s44214-022-00013-x} {\bibfield  {journal} {\bibinfo
  {journal} {Quantum Frontiers}\ }\textbf {\bibinfo {volume} {1}},\ \bibinfo
  {pages} {15} (\bibinfo {year} {2022})}\BibitemShut {NoStop}%
\bibitem [{SM()}]{SM}{See Supplemental Material for additional figures and discussions, which includes Ref. [38].} %
\bibitem [{\citenamefont {M$\rm\ddot{o}$hr-Vorobeva}\ \emph
  {et~al.}(2011)\citenamefont {M$\rm\ddot{o}$hr-Vorobeva}, \citenamefont
  {Johnson}, \citenamefont {Beaud}, \citenamefont {Staub}, \citenamefont
  {De~Souza}, \citenamefont {Milne}, \citenamefont {Ingold}, \citenamefont
  {Demsar}, \citenamefont {Schaefer},\ and\ \citenamefont
  {Titov}}]{MoehrVorobeva2011}%
  \BibitemOpen
  \bibfield  {author} {\bibinfo {author} {\bibfnamefont {E.}~\bibnamefont
  {M\"ohr-Vorobeva}}, \bibinfo {author} {\bibfnamefont {S.~L.}\
  \bibnamefont {Johnson}}, \bibinfo {author} {\bibfnamefont {P.}~\bibnamefont
  {Beaud}}, \bibinfo {author} {\bibfnamefont {U.}~\bibnamefont {Staub}},
  \bibinfo {author} {\bibfnamefont {R.}~\bibnamefont {De~Souza}}, \bibinfo
  {author} {\bibfnamefont {C.}~\bibnamefont {Milne}}, \bibinfo {author}
  {\bibfnamefont {G.}~\bibnamefont {Ingold}}, \bibinfo {author} {\bibfnamefont
  {J.}~\bibnamefont {Demsar}}, \bibinfo {author} {\bibfnamefont
  {H.}~\bibnamefont {Schaefer}},\ and\ \bibinfo {author} {\bibfnamefont
  {A.}~\bibnamefont {Titov}},\ }\href
  {https://doi.org/10.1103/PhysRevLett.107.036403} {\bibfield  {journal}
  {\bibinfo  {journal} {Physical Review Letters}\ }\textbf {\bibinfo {volume}
  {107}},\ \bibinfo {pages} {036403} (\bibinfo {year} {2011})}\BibitemShut
  {NoStop}%
\bibitem [{\citenamefont {Pillo}\ \emph {et~al.}(2000)\citenamefont {Pillo},
  \citenamefont {Hayoz}, \citenamefont {Berger}, \citenamefont
  {L$\rm\acute{e}$vy}, \citenamefont {Schlapbach},\ and\ \citenamefont
  {Aebi}}]{Pillo2000}%
  \BibitemOpen
  \bibfield  {author} {\bibinfo {author} {\bibfnamefont {T.}~\bibnamefont
  {Pillo}}, \bibinfo {author} {\bibfnamefont {J.}~\bibnamefont {Hayoz}},
  \bibinfo {author} {\bibfnamefont {H.}~\bibnamefont {Berger}}, \bibinfo
  {author} {\bibfnamefont {F.}~\bibnamefont {Levy}}, \bibinfo
  {author} {\bibfnamefont {L.}~\bibnamefont {Schlapbach}},\ and\ \bibinfo
  {author} {\bibfnamefont {P.}~\bibnamefont {Aebi}},\ }\href
  {https://doi.org/10.1103/PhysRevB.61.16213} {\bibfield  {journal} {\bibinfo
  {journal} {Physical Review B}\ }\textbf {\bibinfo {volume} {61}},\ \bibinfo
  {pages} {16213} (\bibinfo {year} {2000})}\BibitemShut {NoStop}%
\bibitem [{\citenamefont {Watson}\ \emph {et~al.}(2019)\citenamefont {Watson},
  \citenamefont {Clark}, \citenamefont {Mazzola}, \citenamefont
  {Markovi$\rm\acute{c}$}, \citenamefont {Sunko}, \citenamefont {Kim},
  \citenamefont {Rossnagel},\ and\ \citenamefont {King}}]{Watson2019}%
  \BibitemOpen
  \bibfield  {author} {\bibinfo {author} {\bibfnamefont {M.~D.}\ \bibnamefont
  {Watson}}, \bibinfo {author} {\bibfnamefont {O.~J.}\ \bibnamefont {Clark}},
  \bibinfo {author} {\bibfnamefont {F.}~\bibnamefont {Mazzola}}, \bibinfo
  {author} {\bibfnamefont {I.}~\bibnamefont {Markovi$\rm\acute{c}$}}, \bibinfo
  {author} {\bibfnamefont {V.}~\bibnamefont {Sunko}}, \bibinfo {author}
  {\bibfnamefont {T.~K.}\ \bibnamefont {Kim}}, \bibinfo {author} {\bibfnamefont
  {K.}~\bibnamefont {Rossnagel}},\ and\ \bibinfo {author} {\bibfnamefont
  {P.~D.~C.}\ \bibnamefont {King}},\ }\href
  {https://doi.org/10.1103/PhysRevLett.122.076404} {\bibfield  {journal}
  {\bibinfo  {journal} {Physical Review Letters}\ }\textbf {\bibinfo {volume}
  {122}},\ \bibinfo {pages} {076404} (\bibinfo {year} {2019})}\BibitemShut
  {NoStop}%
\bibitem [{\citenamefont {Yang}\ \emph {et~al.}(2020)\citenamefont {Yang},
  \citenamefont {Rohde}, \citenamefont {Hanff}, \citenamefont {Stange},
  \citenamefont {Xiong}, \citenamefont {Shi}, \citenamefont {Bauer},\ and\
  \citenamefont {Rossnagel}}]{Yang2020}%
  \BibitemOpen
  \bibfield  {author} {\bibinfo {author} {\bibfnamefont {L.~X.}\ \bibnamefont
  {Yang}}, \bibinfo {author} {\bibfnamefont {G.}~\bibnamefont {Rohde}},
  \bibinfo {author} {\bibfnamefont {K.}~\bibnamefont {Hanff}}, \bibinfo
  {author} {\bibfnamefont {A.}~\bibnamefont {Stange}}, \bibinfo {author}
  {\bibfnamefont {R.}~\bibnamefont {Xiong}}, \bibinfo {author} {\bibfnamefont
  {J.}~\bibnamefont {Shi}}, \bibinfo {author} {\bibfnamefont {M.}~\bibnamefont
  {Bauer}},\ and\ \bibinfo {author} {\bibfnamefont {K.}~\bibnamefont
  {Rossnagel}},\ }\href {https://doi.org/10.1103/PhysRevLett.125.266402}
  {\bibfield  {journal} {\bibinfo  {journal} {Phys. Rev. Lett.}\ }\textbf
  {\bibinfo {volume} {125}},\ \bibinfo {pages} {266402} (\bibinfo {year}
  {2020})}\BibitemShut {NoStop}%
\bibitem [{\citenamefont {Chen}\ \emph {et~al.}(2016)\citenamefont {Chen},
  \citenamefont {Chan}, \citenamefont {Fang}, \citenamefont {Mo}, \citenamefont
  {Hussain}, \citenamefont {Fedorov}, \citenamefont {Chou},\ and\ \citenamefont
  {Chiang}}]{Chen2016}%
  \BibitemOpen
  \bibfield  {author} {\bibinfo {author} {\bibfnamefont {P.}~\bibnamefont
  {Chen}}, \bibinfo {author} {\bibfnamefont {Y.-H.}\ \bibnamefont {Chan}},
  \bibinfo {author} {\bibfnamefont {X.-Y.}\ \bibnamefont {Fang}}, \bibinfo
  {author} {\bibfnamefont {S.-K.}\ \bibnamefont {Mo}}, \bibinfo {author}
  {\bibfnamefont {Z.}~\bibnamefont {Hussain}}, \bibinfo {author} {\bibfnamefont
  {A.-V.}\ \bibnamefont {Fedorov}}, \bibinfo {author} {\bibfnamefont {M.~Y.}\
  \bibnamefont {Chou}},\ and\ \bibinfo {author} {\bibfnamefont {T.-C.}\
  \bibnamefont {Chiang}},\ }\href {https://doi.org/10.1038/srep37910}
  {\bibfield  {journal} {\bibinfo  {journal} {Scientific Reports}\ }\textbf
  {\bibinfo {volume} {6}},\ \bibinfo {pages} {37910} (\bibinfo {year}
  {2016})}\BibitemShut {NoStop}%
\bibitem [{\citenamefont {Monney}\ \emph {et~al.}(2010)\citenamefont {Monney},
  \citenamefont {Schwier}, \citenamefont {Garnier}, \citenamefont {Mariotti},
  \citenamefont {Didiot}, \citenamefont {Beck}, \citenamefont {Aebi},
  \citenamefont {Cercellier}, \citenamefont {Marcus}, \citenamefont
  {Battaglia}, \citenamefont {Berger},\ and\ \citenamefont
  {Titov}}]{Monney2010}%
  \BibitemOpen
  \bibfield  {author} {\bibinfo {author} {\bibfnamefont {C.}~\bibnamefont
  {Monney}}, \bibinfo {author} {\bibfnamefont {E.~F.}\ \bibnamefont {Schwier}},
  \bibinfo {author} {\bibfnamefont {M.~G.}\ \bibnamefont {Garnier}}, \bibinfo
  {author} {\bibfnamefont {N.}~\bibnamefont {Mariotti}}, \bibinfo {author}
  {\bibfnamefont {C.}~\bibnamefont {Didiot}}, \bibinfo {author} {\bibfnamefont
  {H.}~\bibnamefont {Beck}}, \bibinfo {author} {\bibfnamefont {P.}~\bibnamefont
  {Aebi}}, \bibinfo {author} {\bibfnamefont {H.}~\bibnamefont {Cercellier}},
  \bibinfo {author} {\bibfnamefont {J.}~\bibnamefont {Marcus}}, \bibinfo
  {author} {\bibfnamefont {C.}~\bibnamefont {Battaglia}}, \bibinfo {author}
  {\bibfnamefont {H.}~\bibnamefont {Berger}},\ and\ \bibinfo {author}
  {\bibfnamefont {A.~N.}\ \bibnamefont {Titov}},\ }\href
  {https://doi.org/10.1103/PhysRevB.81.155104} {\bibfield  {journal} {\bibinfo
  {journal} {Physical Review B}\ }\textbf {\bibinfo {volume} {81}},\ \bibinfo
  {pages} {155104} (\bibinfo {year} {2010})}\BibitemShut {NoStop}%
\bibitem [{\citenamefont {Trigo}\ \emph {et~al.}(2021)\citenamefont {Trigo},
  \citenamefont {Giraldo-Gallo}, \citenamefont {Clark}, \citenamefont {Kozina},
  \citenamefont {Henighan}, \citenamefont {Jiang}, \citenamefont {Chollet},
  \citenamefont {Fisher}, \citenamefont {Glownia}, \citenamefont {Katayama},
  \citenamefont {Kirchmann}, \citenamefont {Leuenberger}, \citenamefont {Liu},
  \citenamefont {Reis}, \citenamefont {Shen},\ and\ \citenamefont
  {Zhu}}]{Trigo2021}%
  \BibitemOpen
  \bibfield  {author} {\bibinfo {author} {\bibfnamefont {M.}~\bibnamefont
  {Trigo}}, \bibinfo {author} {\bibfnamefont {P.}~\bibnamefont
  {Giraldo-Gallo}}, \bibinfo {author} {\bibfnamefont {J.~N.}\ \bibnamefont
  {Clark}}, \bibinfo {author} {\bibfnamefont {M.~E.}\ \bibnamefont {Kozina}},
  \bibinfo {author} {\bibfnamefont {T.}~\bibnamefont {Henighan}}, \bibinfo
  {author} {\bibfnamefont {M.~P.}\ \bibnamefont {Jiang}}, \bibinfo {author}
  {\bibfnamefont {M.}~\bibnamefont {Chollet}}, \bibinfo {author} {\bibfnamefont
  {I.~R.}\ \bibnamefont {Fisher}}, \bibinfo {author} {\bibfnamefont {J.~M.}\
  \bibnamefont {Glownia}}, \bibinfo {author} {\bibfnamefont {T.}~\bibnamefont
  {Katayama}}, \bibinfo {author} {\bibfnamefont {P.~S.}\ \bibnamefont
  {Kirchmann}}, \bibinfo {author} {\bibfnamefont {D.}~\bibnamefont
  {Leuenberger}}, \bibinfo {author} {\bibfnamefont {H.}~\bibnamefont {Liu}},
  \bibinfo {author} {\bibfnamefont {D.~A.}\ \bibnamefont {Reis}}, \bibinfo
  {author} {\bibfnamefont {Z.~X.}\ \bibnamefont {Shen}},\ and\ \bibinfo
  {author} {\bibfnamefont {D.}~\bibnamefont {Zhu}},\ }\href
  {https://doi.org/10.1103/PhysRevB.103.054109} {\bibfield  {journal} {\bibinfo
   {journal} {Physical Review B}\ }\textbf {\bibinfo {volume} {103}},\ \bibinfo
  {pages} {054109} (\bibinfo {year} {2021})}\BibitemShut {NoStop}%
\bibitem [{\citenamefont {Yusupov}\ \emph {et~al.}(2010)\citenamefont
  {Yusupov}, \citenamefont {Mertelj}, \citenamefont {Kabanov}, \citenamefont
  {Brazovskii}, \citenamefont {Kusar}, \citenamefont {Chu}, \citenamefont
  {Fisher},\ and\ \citenamefont {Mihailovic}}]{Yusupov2010}%
  \BibitemOpen
  \bibfield  {author} {\bibinfo {author} {\bibfnamefont {R.}~\bibnamefont
  {Yusupov}}, \bibinfo {author} {\bibfnamefont {T.}~\bibnamefont {Mertelj}},
  \bibinfo {author} {\bibfnamefont {V.~V.}\ \bibnamefont {Kabanov}}, \bibinfo
  {author} {\bibfnamefont {S.}~\bibnamefont {Brazovskii}}, \bibinfo {author}
  {\bibfnamefont {P.}~\bibnamefont {Kusar}}, \bibinfo {author} {\bibfnamefont
  {J.-H.}\ \bibnamefont {Chu}}, \bibinfo {author} {\bibfnamefont {I.~R.}\
  \bibnamefont {Fisher}},\ and\ \bibinfo {author} {\bibfnamefont
  {D.}~\bibnamefont {Mihailovic}},\ }\href {https://doi.org/10.1038/nphys1738}
  {\bibfield  {journal} {\bibinfo  {journal} {Nature Physics}\ }\textbf
  {\bibinfo {volume} {6}},\ \bibinfo {pages} {681} (\bibinfo {year}
  {2010})}\BibitemShut {NoStop}%
\bibitem [{\citenamefont {Heinrich}\ \emph {et~al.}(2022)\citenamefont
  {Heinrich}, \citenamefont {Chang}, \citenamefont {Zayko}, \citenamefont
  {Rossnagel}, \citenamefont {Sivis},\ and\ \citenamefont {Ropers}}]{heinrich}%
  \BibitemOpen
  \bibfield  {author} {\bibinfo {author} {\bibfnamefont {T.}~\bibnamefont
  {Heinrich}}, \bibinfo {author} {\bibfnamefont {H.-T.}\ \bibnamefont {Chang}},
  \bibinfo {author} {\bibfnamefont {S.}~\bibnamefont {Zayko}}, \bibinfo
  {author} {\bibfnamefont {K.}~\bibnamefont {Rossnagel}}, \bibinfo {author}
  {\bibfnamefont {M.}~\bibnamefont {Sivis}},\ and\ \bibinfo {author}
  {\bibfnamefont {C.}~\bibnamefont {Ropers}},\ }\href@noop {} {\bibfield
  {journal} {\bibinfo  {journal} {arXiv:2211.03562}\ } (\bibinfo {year}
  {2022})}\BibitemShut {NoStop}%
\bibitem [{\citenamefont {Lian}\ \emph {et~al.}(2020)\citenamefont {Lian},
  \citenamefont {Zhang}, \citenamefont {Hu}, \citenamefont {Guan},\ and\
  \citenamefont {Meng}}]{Lian2020}%
  \BibitemOpen
  \bibfield  {author} {\bibinfo {author} {\bibfnamefont {C.}~\bibnamefont
  {Lian}}, \bibinfo {author} {\bibfnamefont {S.-J.}\ \bibnamefont {Zhang}},
  \bibinfo {author} {\bibfnamefont {S.-Q.}\ \bibnamefont {Hu}}, \bibinfo
  {author} {\bibfnamefont {M.-X.}\ \bibnamefont {Guan}},\ and\ \bibinfo
  {author} {\bibfnamefont {S.}~\bibnamefont {Meng}},\ }\href
  {https://doi.org/10.1038/s41467-019-13672-7} {\bibfield  {journal} {\bibinfo
  {journal} {Nature Communications}\ }\textbf {\bibinfo {volume} {11}},\
  \bibinfo {pages} {43} (\bibinfo {year} {2020})}\BibitemShut {NoStop}%
\bibitem [{\citenamefont {Okazaki}\ \emph {et~al.}(2018)\citenamefont
  {Okazaki}, \citenamefont {Ogawa}, \citenamefont {Suzuki}, \citenamefont
  {Yamamoto}, \citenamefont {Someya}, \citenamefont {Michimae}, \citenamefont
  {Watanabe}, \citenamefont {Lu}, \citenamefont {Nohara}, \citenamefont
  {Takagi}, \citenamefont {Katayama}, \citenamefont {Sawa}, \citenamefont
  {Fujisawa}, \citenamefont {Kanai}, \citenamefont {Ishii}, \citenamefont
  {Itatani}, \citenamefont {Mizokawa},\ and\ \citenamefont
  {Shin}}]{Okazaki2018}%
  \BibitemOpen
  \bibfield  {author} {\bibinfo {author} {\bibfnamefont {K.}~\bibnamefont
  {Okazaki}}, \bibinfo {author} {\bibfnamefont {Y.}~\bibnamefont {Ogawa}},
  \bibinfo {author} {\bibfnamefont {T.}~\bibnamefont {Suzuki}}, \bibinfo
  {author} {\bibfnamefont {T.}~\bibnamefont {Yamamoto}}, \bibinfo {author}
  {\bibfnamefont {T.}~\bibnamefont {Someya}}, \bibinfo {author} {\bibfnamefont
  {S.}~\bibnamefont {Michimae}}, \bibinfo {author} {\bibfnamefont
  {M.}~\bibnamefont {Watanabe}}, \bibinfo {author} {\bibfnamefont
  {Y.}~\bibnamefont {Lu}}, \bibinfo {author} {\bibfnamefont {M.}~\bibnamefont
  {Nohara}}, \bibinfo {author} {\bibfnamefont {H.}~\bibnamefont {Takagi}},
  \bibinfo {author} {\bibfnamefont {N.}~\bibnamefont {Katayama}}, \bibinfo
  {author} {\bibfnamefont {H.}~\bibnamefont {Sawa}}, \bibinfo {author}
  {\bibfnamefont {M.}~\bibnamefont {Fujisawa}}, \bibinfo {author}
  {\bibfnamefont {T.}~\bibnamefont {Kanai}}, \bibinfo {author} {\bibfnamefont
  {N.}~\bibnamefont {Ishii}}, \bibinfo {author} {\bibfnamefont
  {J.}~\bibnamefont {Itatani}}, \bibinfo {author} {\bibfnamefont
  {T.}~\bibnamefont {Mizokawa}},\ and\ \bibinfo {author} {\bibfnamefont
  {S.}~\bibnamefont {Shin}},\ }\href
  {https://doi.org/10.1038/s41467-018-06801-1} {\bibfield  {journal} {\bibinfo
  {journal} {Nature Communications}\ }\textbf {\bibinfo {volume} {9}},\
  \bibinfo {pages} {4322} (\bibinfo {year} {2018})}\BibitemShut {NoStop}%
\bibitem [{\citenamefont {Mor}\ \emph {et~al.}(2017)\citenamefont {Mor},
  \citenamefont {Herzog}, \citenamefont {Gole\ifmmode~\check{z}\else
  \v{z}\fi{}}, \citenamefont {Werner}, \citenamefont {Eckstein}, \citenamefont
  {Katayama}, \citenamefont {Nohara}, \citenamefont {Takagi}, \citenamefont
  {Mizokawa}, \citenamefont {Monney},\ and\ \citenamefont
  {St\"ahler}}]{Mor2017}%
  \BibitemOpen
  \bibfield  {author} {\bibinfo {author} {\bibfnamefont {S.}~\bibnamefont
  {Mor}}, \bibinfo {author} {\bibfnamefont {M.}~\bibnamefont {Herzog}},
  \bibinfo {author} {\bibfnamefont {D.}~\bibnamefont
  {Gole\ifmmode~\check{z}\else \v{z}\fi{}}}, \bibinfo {author} {\bibfnamefont
  {P.}~\bibnamefont {Werner}}, \bibinfo {author} {\bibfnamefont
  {M.}~\bibnamefont {Eckstein}}, \bibinfo {author} {\bibfnamefont
  {N.}~\bibnamefont {Katayama}}, \bibinfo {author} {\bibfnamefont
  {M.}~\bibnamefont {Nohara}}, \bibinfo {author} {\bibfnamefont
  {H.}~\bibnamefont {Takagi}}, \bibinfo {author} {\bibfnamefont
  {T.}~\bibnamefont {Mizokawa}}, \bibinfo {author} {\bibfnamefont
  {C.}~\bibnamefont {Monney}},\ and\ \bibinfo {author} {\bibfnamefont
  {J.}~\bibnamefont {St\"ahler}},\ }\href
  {https://doi.org/10.1103/PhysRevLett.119.086401} {\bibfield  {journal}
  {\bibinfo  {journal} {Physical Review Letters}\ }\textbf {\bibinfo {volume}
  {119}},\ \bibinfo {pages} {086401} (\bibinfo {year} {2017})}\BibitemShut
  {NoStop}%
\bibitem [{\citenamefont {Rohwer}\ \emph {et~al.}(2011)\citenamefont {Rohwer},
  \citenamefont {Hellmann}, \citenamefont {Wiesenmayer}, \citenamefont {Sohrt},
  \citenamefont {Stange}, \citenamefont {Slomski}, \citenamefont {Carr},
  \citenamefont {Liu}, \citenamefont {Miaja-Avila}, \citenamefont {Kallaene},
  \citenamefont {Mathias}, \citenamefont {Kipp}, \citenamefont {Rossnagel},\
  and\ \citenamefont {Bauer}}]{Rohwer2011}%
  \BibitemOpen
  \bibfield  {author} {\bibinfo {author} {\bibfnamefont {T.}~\bibnamefont
  {Rohwer}}, \bibinfo {author} {\bibfnamefont {S.}~\bibnamefont {Hellmann}},
  \bibinfo {author} {\bibfnamefont {M.}~\bibnamefont {Wiesenmayer}}, \bibinfo
  {author} {\bibfnamefont {C.}~\bibnamefont {Sohrt}}, \bibinfo {author}
  {\bibfnamefont {A.}~\bibnamefont {Stange}}, \bibinfo {author} {\bibfnamefont
  {B.}~\bibnamefont {Slomski}}, \bibinfo {author} {\bibfnamefont
  {A.}~\bibnamefont {Carr}}, \bibinfo {author} {\bibfnamefont {Y.}~\bibnamefont
  {Liu}}, \bibinfo {author} {\bibfnamefont {L.}~\bibnamefont {Miaja-Avila}},
  \bibinfo {author} {\bibfnamefont {M.}~\bibnamefont {Kallaene}}, \bibinfo
  {author} {\bibfnamefont {S.}~\bibnamefont {Mathias}}, \bibinfo {author}
  {\bibfnamefont {L.}~\bibnamefont {Kipp}}, \bibinfo {author} {\bibfnamefont
  {K.}~\bibnamefont {Rossnagel}},\ and\ \bibinfo {author} {\bibfnamefont
  {M.}~\bibnamefont {Bauer}},\ }\href {https://doi.org/10.1038/nature09829}
  {\bibfield  {journal} {\bibinfo  {journal} {Nature}\ }\textbf {\bibinfo
  {volume} {471}},\ \bibinfo {pages} {490} (\bibinfo {year}
  {2011})}\BibitemShut{NoStop}%
\end{thebibliography}

%

\newpage

\begin{figure*}
\includegraphics[width=2.1\columnwidth,page=1]{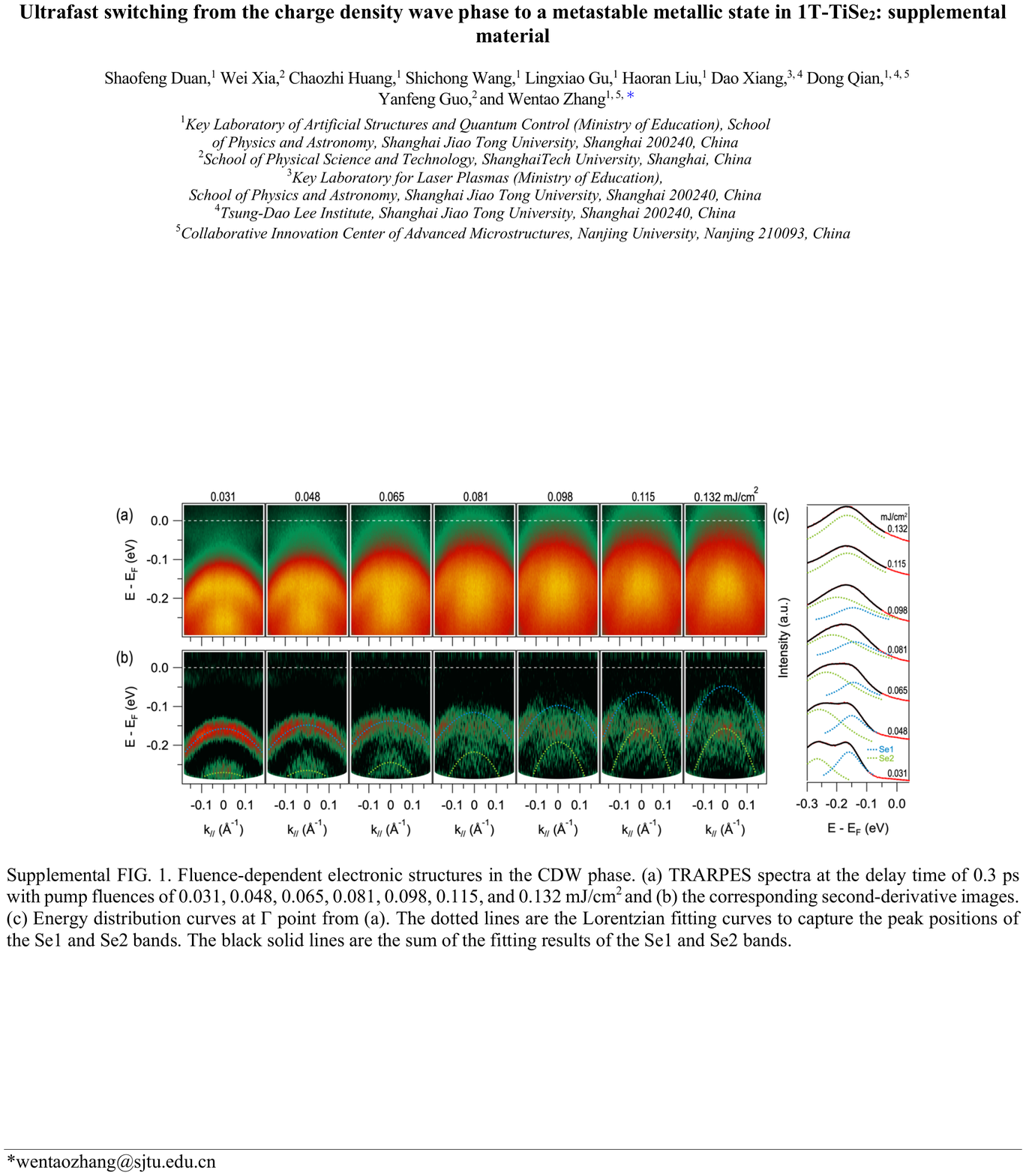}
\end{figure*}
\begin{figure*}
\includegraphics[width=2.1\columnwidth,page=2]{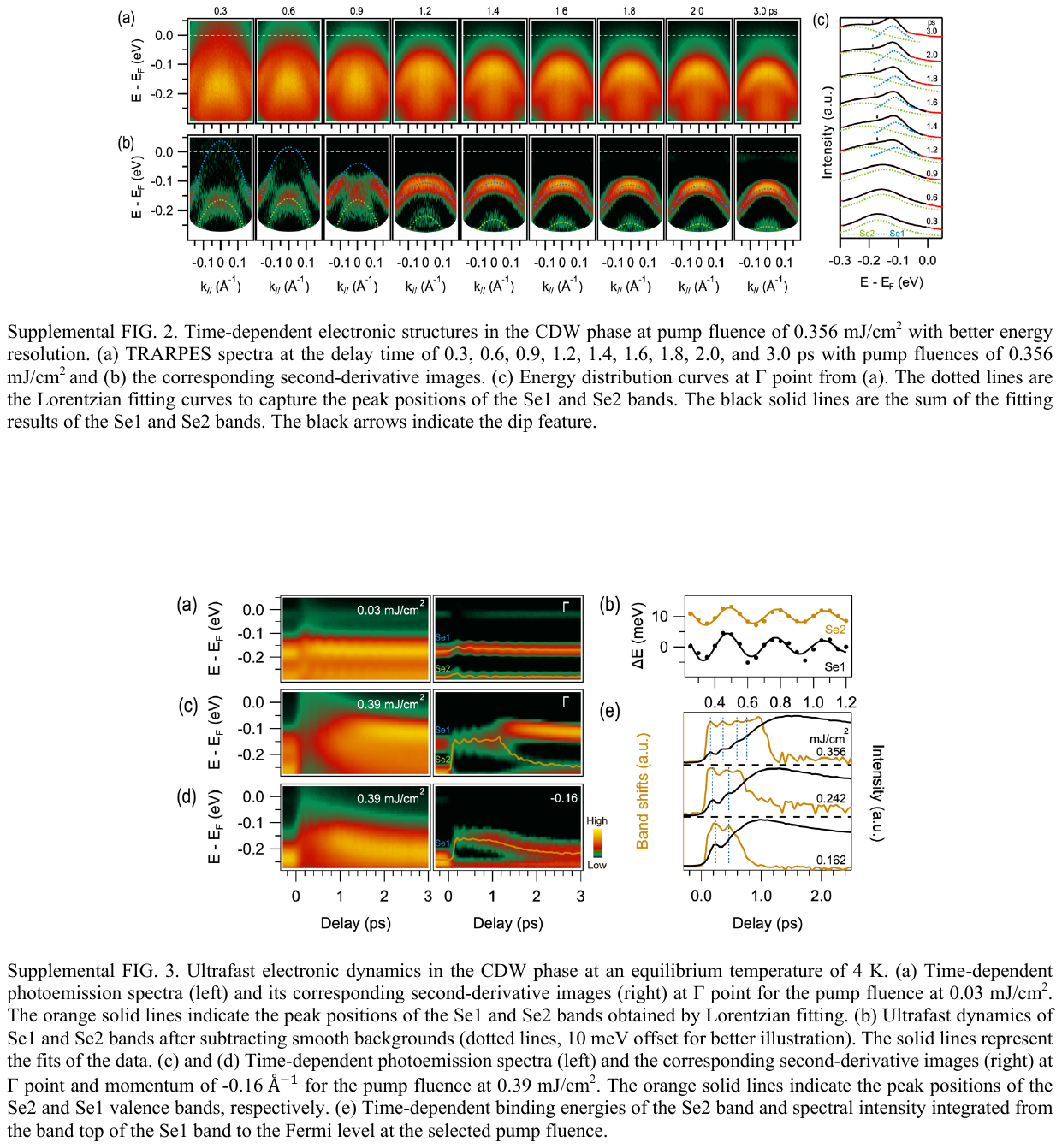}
\end{figure*}
\begin{figure*}
\includegraphics[width=2.1\columnwidth,page=3]{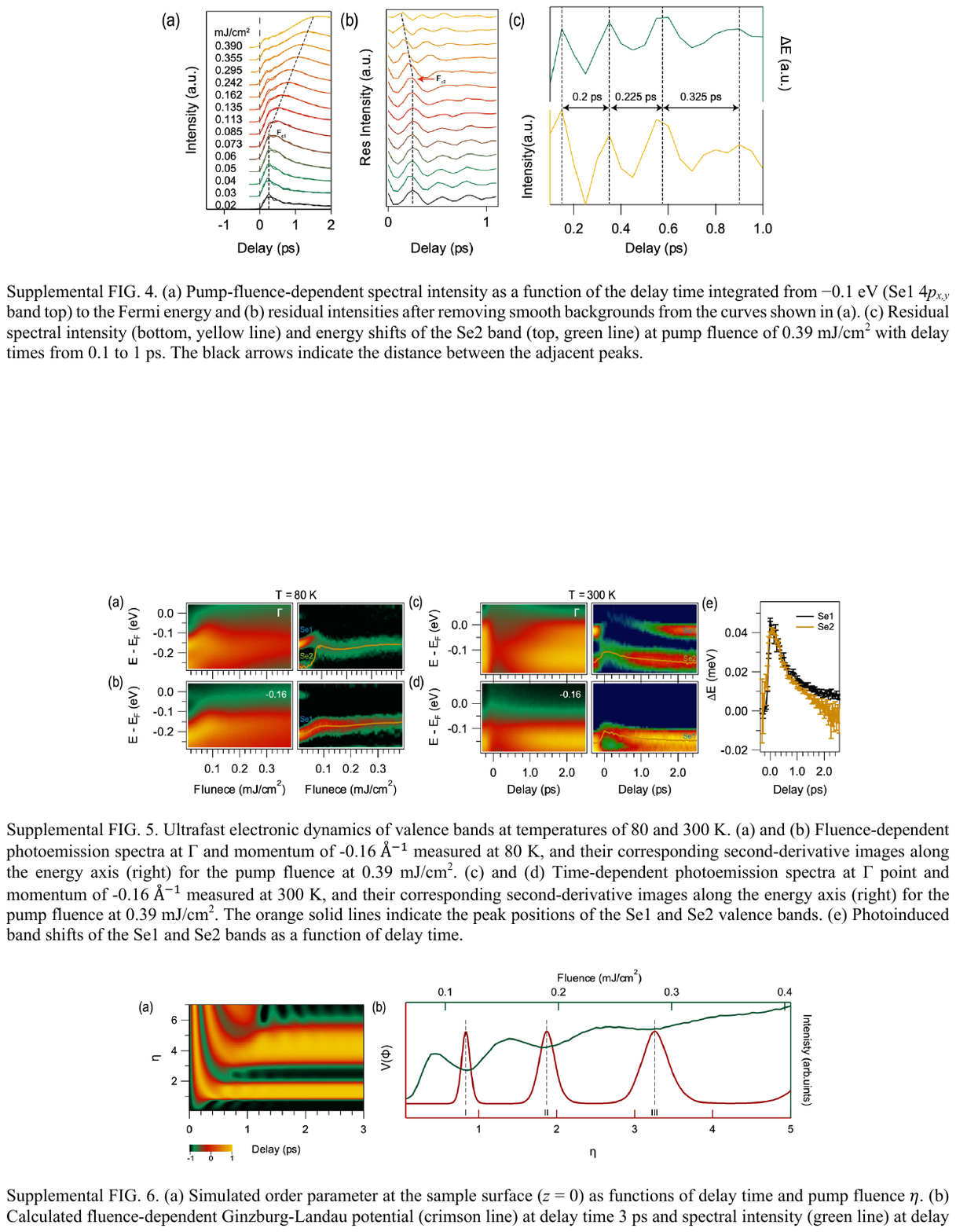}
\end{figure*}
\begin{figure*}
\includegraphics[width=2.1\columnwidth,page=4]{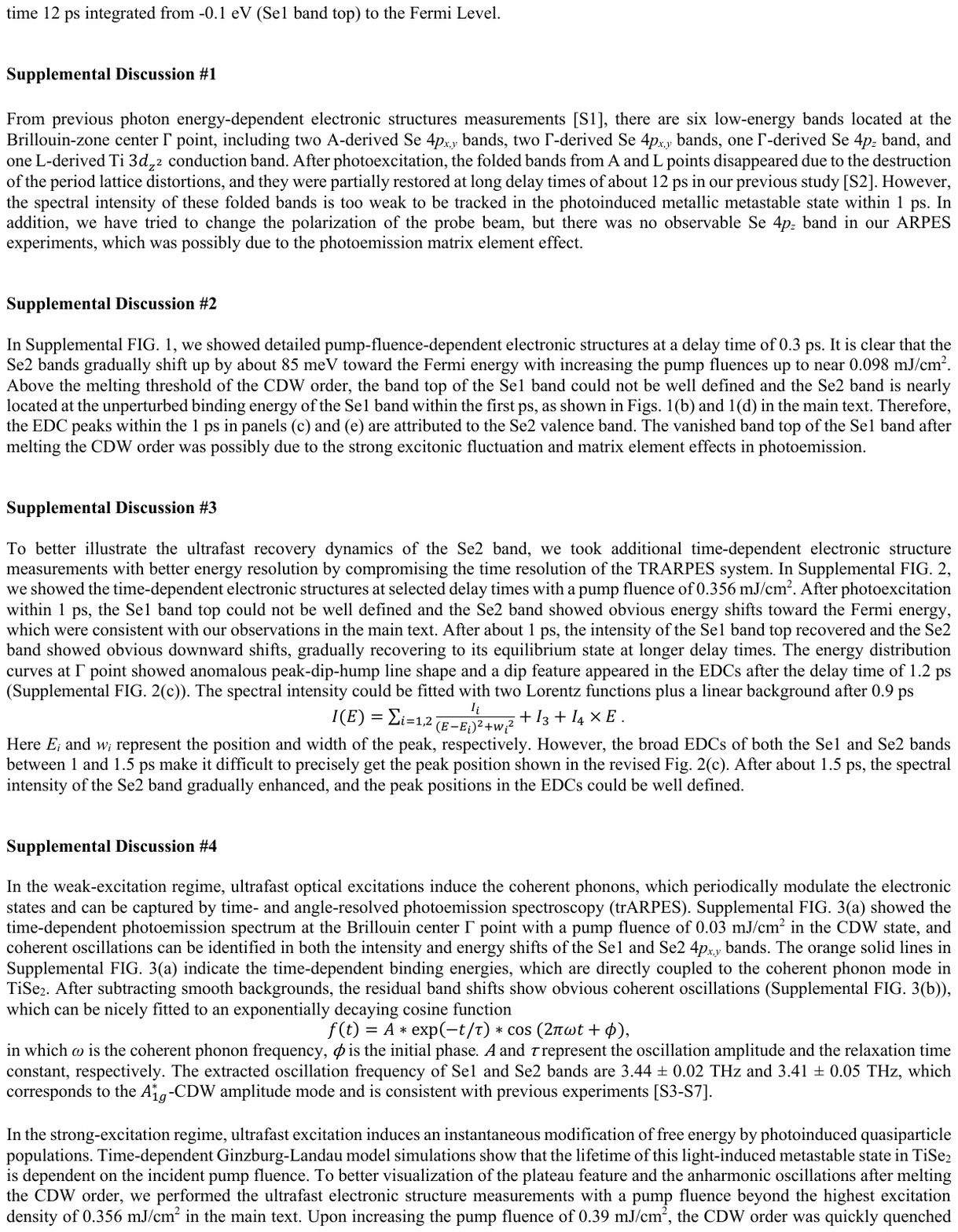}
\end{figure*}
\begin{figure*}
\includegraphics[width=2.1\columnwidth,page=5]{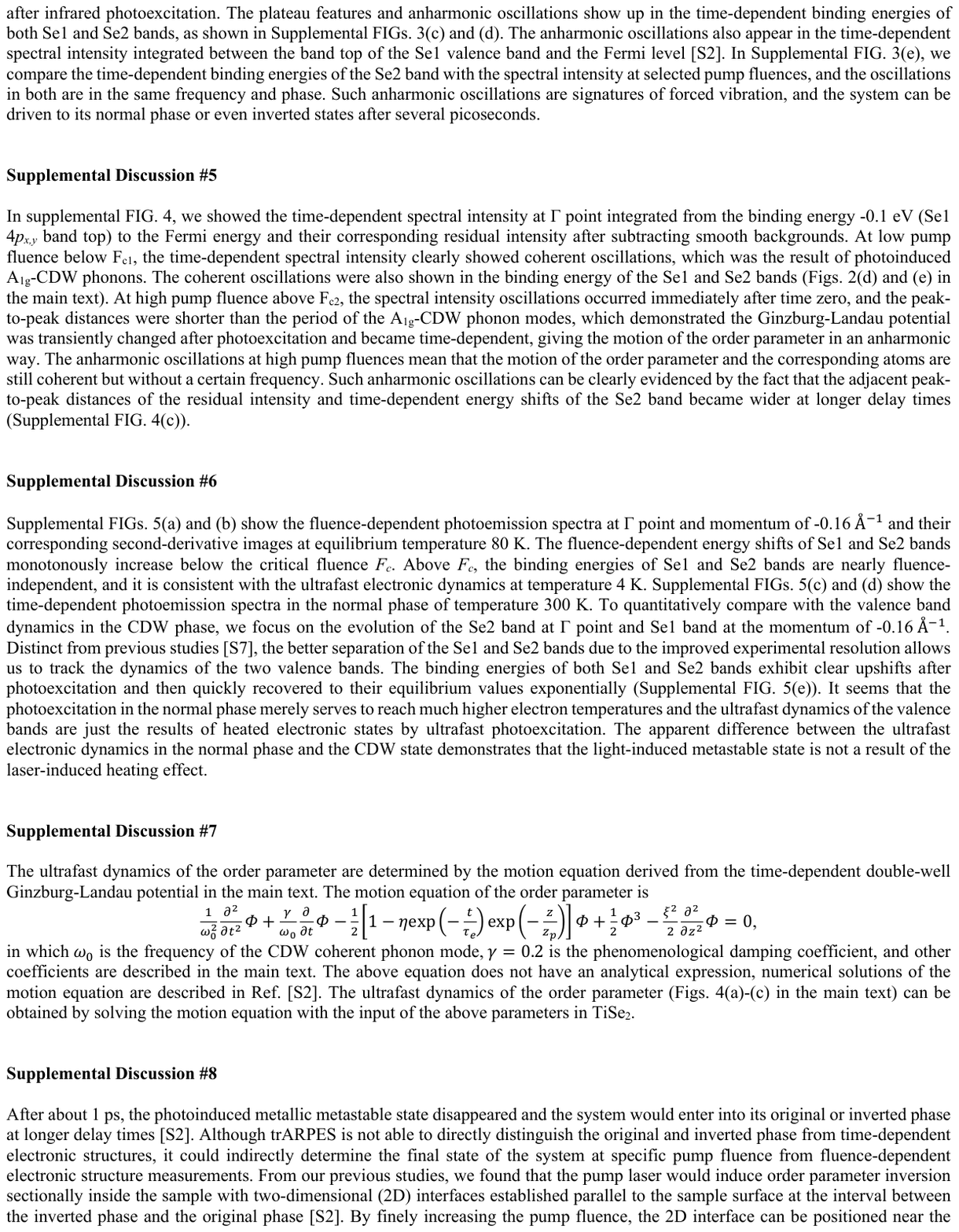}
\end{figure*}
\begin{figure*}
\includegraphics[width=2.1\columnwidth,page=6]{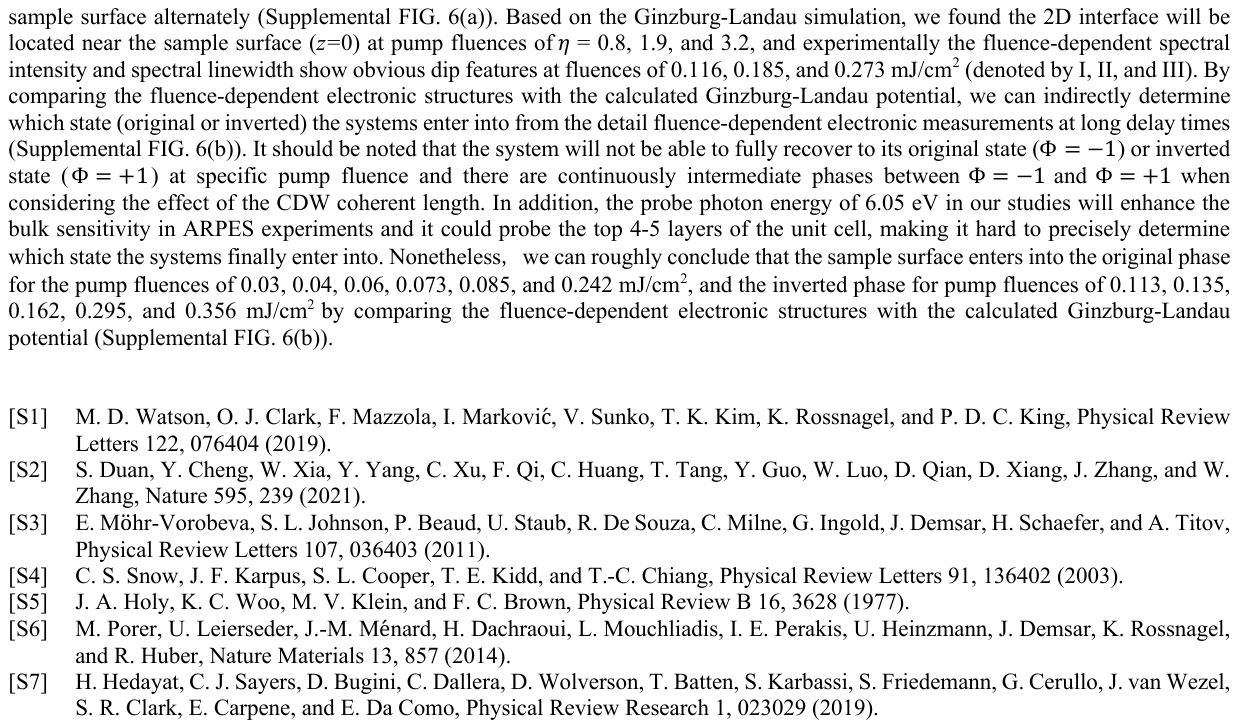}
\end{figure*}

\end{document}